\documentclass[10pt,twocolumn,twoside]{IEEEtran}
\usepackage{array}
\usepackage[table, xcdraw, dvipsnames]{xcolor}
\usepackage{amsmath, amsfonts, bbm, breqn, mathtools, mathrsfs, bm, amssymb}
\usepackage[export]{adjustbox}
\usepackage{subcaption, caption, makecell, multirow, graphicx}
\usepackage{float}
\usepackage{textcomp}
\usepackage{changes, nameref}
\usepackage{txfonts}
\usepackage{url}
\usepackage{verbatim}
\usepackage{cite}
\usepackage{booktabs} 
\newcommand\numberthis{\addtocounter{equation}{1}\tag{\theequation}}

\usepackage{algpseudocode} 
\usepackage{soul}
\usepackage[colorlinks, 
            linkcolor=red,      
            anchorcolor=red,  
            citecolor=blue,       
            ]{hyperref}

\begin{document}

    \title{Towards Practical Single-shot  Phase Retrieval with Physics-Driven Deep Neural Network}
    
    \author{\IEEEauthorblockN{Qiuliang Ye\IEEEauthorrefmark{1}\IEEEauthorrefmark{2}, Li-Wen Wang\IEEEauthorrefmark{2}, and Daniel Pak-Kong Lun\IEEEauthorrefmark{1},~\IEEEmembership{Senior Member,~IEEE}
    }
    \thanks{Manuscript received ?? ??, 2022; revised ??? ??, 2022 and ??? ??, 2022; accepted ??? ??, 2022. Date of publication ??? ??, 2022; date of current version ??? ??, 2022. The work presented in this article was supported by the Hong Kong Research Grant Council under General Research Fund no. PolyU 15225321. The associate editor coordinating the review of this manuscript and approving it for publication was ????? ????? ?????. (Corresponding author: Qiuliang Ye and Daniel Pak-Kong Lun.)
    }
    \thanks{Qiuliang Ye, Li-Wen Wang, and Daniel Pak-Kong Lun are with the Department of Electronic and Information Engineering, The Hong Kong Polytechnic University, Kowloon, Hong Kong SAR, China. Daniel Pak-Kong Lun is also with Photonics Research Institute, The Hong Kong Polytechnic University, Kowloon, Hong Kong SAR, China. Email: \href{mailto:qiu-liang.ye@connect.polyu.hk}{qiu-liang.ye@connect.polyu.hk} and  \href{mailto:pak.kong.lun@polyu.edu.hk}{pak.kong.lun@polyu.edu.hk}.}
    \thanks{ \IEEEauthorrefmark{2} These authors contribute equally to this work as first authors. }
    \thanks{Digital Object Identifier 10.1109/TIP.2022.???????}}
    
    \markboth{IEEE Transactions on Image Processing}%
    {Ye \MakeLowercase{\textit{et al.}}: Towards Practical Single-shot  Phase Retrieval with Physics-Driven Deep Neural Network}

    \maketitle
    
    \begin{abstract}
    	
    	Phase retrieval (PR), a long-established challenge for recovering a complex-valued signal from its Fourier intensity-only measurements, has attracted considerable attention due to its widespread applications in digital imaging.  Recently, deep learning-based approaches were developed that achieved some success in single-shot PR. These approaches require a single Fourier intensity measurement without the need to impose any additional constraints on the measured data. Nevertheless, vanilla deep neural networks (DNN) do not give good performance due to the substantial disparity between the input and output domains of the PR problems. Physics-informed approaches try to incorporate the Fourier intensity measurements into an iterative approach to increase the reconstruction accuracy. It, however, requires a lengthy computation process, and the accuracy still cannot be guaranteed. Besides, many of these approaches work on simulation data that ignore some common problems such as saturation and quantization errors in practical optical PR systems. In this paper, a novel physics-driven multi-scale DNN structure dubbed \textit{PPRNet} is proposed. Similar to other deep learning-based PR methods, \textit{PPRNet} requires only a single Fourier intensity measurement. It is physics-driven that the network is guided to follow the Fourier intensity measurement at different scales to enhance the reconstruction accuracy. \textit{PPRNet} has a feedforward structure and can be end-to-end trained. Thus, it is much faster and more accurate than the traditional physics-driven PR approaches. Extensive simulations and experiments on a practical optical platform were conducted. The results demonstrate the superiority and practicality of the proposed \textit{PPRNet} over the traditional learning-based PR methods.
    \end{abstract}
    

    \section{Introduction} \label{Sec:Introdcution}
    
    \IEEEPARstart{P}{hase} retrieval (PR) intends to reconstruct a complex-valued signal only from its Fourier intensity measurements. It is a key problem in crystallography \cite{Gerchberg1972APA, Millane:90}, optical imaging \cite{Fienup:82}, astronomical imaging \cite{fienup1987phase}, diffraction imaging \cite{rodenburg2008ptychography}, etc. PR is also a crucial component of holographic imaging \cite{Loop_PR_2020}. The investigation of PR methods was initiated in the 1970s. Numerous reconstruction approaches were developed by the optics research community \cite{Gerchberg1972APA, Fienup:82, Luke_2004}. Recently, the developments in modern optimization theories \cite{candes2015phase, candes2015code, Chen_15} and computational imaging \cite{Katz_2014, Shechtman2015PhaseRW} provided further understanding of the problem. From the mathematical perspective, the PR problem can be expressed as follows \cite{candes2015code}:
    \begin{equation}
    	\label{Eq:CDP}
    	\begin{aligned}
    		\text { Find } \mathbf{x} \in \mathbb{C}^{N} \quad \text { s.t. } \ \bm{\mathcal{X}}_{m}=\left|\mathcal{F}\left(\mathbf{h}_{m} \circ \mathbf{x}\right)\right|^{2}, m=1, \ldots, M,
    	\end{aligned}
    \end{equation}
    where $\mathbf{x}\in \mathbb{C}^N$ is the complex-valued signal of interest; $ \bm{\mathcal{X}}$ denotes the Fourier intensity measurements; $\circ$ and $\mathcal{F}$ stand for elementwise multiplication and Fourier transform, respectively. The pre-determined optical masks $ \mathbf{h} $ are optional; they are for providing the constraints to lessen the problem’s ill-posedness. There are various methods for the implementation of the masks. For example, early PR approaches considered the non-zero signal support as the optical masks \cite{Fienup:82}. However, these traditional methods cannot ensure globally optimal solutions (uniqueness condition) nor the convergence of the optimization procedure. In recent years, random masks were employed as powerful constraints for the optimization process \cite{candes2015phase, YE2022106808}. They can be implemented using a digital micromirror device (DMD) or spatial light modulator (SLM) \cite{Horisaki2014SingleshotPI, Zheng2017DigitalMD}. Although the use of random masks can improve the reconstruction performance, the high cost and inaccuracy of the DMD and SLM devices deter the general application of the method. Besides, it is empirically shown in \cite{candes2015code, candes2015phase} that around $ 4 - 6 $  measurements are needed for accurate reconstruction with random binary masks. It increases the data acquisition time and is thus undesirable for dynamic applications. For solving \eqref{Eq:CDP}, different iterative optimization methods, such as ADMM \cite{chang2018total}, Wirtinger Flow \cite{candes2015phase}, etc., are generally used. These approaches are extremely time-consuming. Thus, the resulting PR methods are not suitable for any real-time applications. 
    
    In the last three decades, deep neural networks (DNN) have been widely studied and successfully applied to different applications \cite{LeCun_2015}. They were also used in the PR problems \cite{Sinha17, Shi:19, deng_li_goy_kang_barbastathis_2020, Zhang_21, pmlr_metzler18a, uelwer2021phase, Uelwer_PhaseRetrieval, Wu_cw5029, DeepPhaseCut_2021, I_l_2019}. Different from the traditional optimization-based approaches, the deep learning-based PR methods can work with only a single Fourier intensity measurement. These methods can fit in some feedforward DNN structures to achieve real-time performance when processing with GPU. Nevertheless, the accuracy of these methods still has much room to improve. It is still a challenge to directly infer a complex-valued signal from its Fourier intensity due to the enormous discrepancy between a signal in the Fourier and spatial domains. To improve the accuracy, researchers also suggested the physics-driven method \cite{pmlr_metzler18a} in which the intensity measurement was used to inform a plug-and-play optimization process. However, the method is iterative and as time-consuming as the traditional optimization-based methods. Besides using the plug-and-play structure, researchers also introduced the physics information to the PR process by directly plugging in the HIO algorithm \cite{Fienup:82} running alternately with a DNN in an iterative procedure \cite{I_l_2019}. These physics-driven approaches have a common characteristic that they apply the physics information to a traditional optimization method and let it work iteratively with a network model. However, in this case, the system cannot be end-to-end trained. The estimation error of the optimization algorithm may have a special distribution that is unknown to the network model. The iterative process can thus be trapped at a local minimum and fails to give the best solution. 
    
    One common problem of these deep learning-based PR methods is that most of them are trained and tested with simulation data. Except for noises, they often ignored the other artifacts in practical Fourier intensity measurements. It makes their reported results unreliable. Most PR applications involve structured images, which have energy concentrated at low frequencies (in particular, the d.c.). The dynamic range of the data in an intensity measurement is thus extremely large. Most general imaging devices nowadays have only a $12$ to $16$ bits dynamic range. It makes the low-frequency part of the intensity image severely saturated. As we have shown in Section \ref{sec:PRsystem}, the saturation problem can significantly affect the performance of PR methods.       
    
    To rectify the abovementioned problems, we propose in this paper a novel physics-driven deep learning-based PR method dubbed \textit{PPRNet}. Similar to other deep learning-based PR methods, \textit{PPRNet} requires only one Fourier intensity measurement for each PR reconstruction. It, however, gives a much higher accuracy by using a physics-driven method that guides the network to follow the Fourier intensity measurement at different scales to reconstruct the images. The resulting network structure is still of feedforward type. Thus, it is much faster than the iterative physics-driven approaches. The whole network can be end-to-end trained to give the best result. \textit{PPRNet} is enabled by a novel Hybrid Unwinding Block (HUB) embedded in a multi-scale convolutional neural network (CNN) structure. It directs the input feature map into two paths such that the global and local information of the feature map at different scales are separately processed with physics informed. The data are then combined with a channel attention method so that the significant features are collected for reconstructing the images.  To evaluate the generality of the proposed \textit{PPRNet}, we have conducted a series of simulations with $2$ different datasets that contain complex-valued images with linearly correlated and uncorrelated magnitude and phase. All data in the intensity image were capped to be represented by $12$-bit integers to simulate the saturation problem with quantization errors. We then adopted the defocusing method \cite{Ye_SiSPRNet} to mitigate the saturation problem in the intensity images. As shown in the simulation results, the proposed \textit{PPRNet} significantly outperforms the state-of-the-art deep learning-based PR methods. To understand the performance of the proposed \textit{PPRNet} in practical applications, we constructed an optical setup for generating the intensity measurements of phase-only images obtained from $3$ datasets. Naturally, all intensity measurements had a serious saturation problem in their low-frequency data. We again used the defocusing method to mitigate the effect of the problem. They were then used for the training and testing of the proposed \textit{PPRNet}. Based on these experimental data, we compared the proposed \textit{PPRNet} with the state-of-the-art deep learning-based PR methods. A significant improvement in the accuracy is noted. It also has lower complexity than other physics-driven PR methods.
    
    To summarize, the contribution of this work is three-folded:
    
    \noindent\textbf{1.} A physics-driven deep learning-based PR method, namely \textit{PPRNet}, is developed. It requires only a single Fourier intensity measurement for each PR reconstruction without the need for any additional masking scheme to impose constraints on the measurement. It allows the Fourier intensity measurement to inform the training and inferencing processes so that the model is guided to give the right solution. Experimental results have demonstrated the effectiveness of this approach and the improvement it brings over the traditional deep learning-based PR methods. \textit{PPRNet} has a non-iterative feedforward structure and is end-to-end trained. It has lower complexity than the existing physics-driven approaches while having higher accuracy. 
    
    \noindent\textbf{2.} A novel Hybrid Unwinding Block (HUB) is proposed. While the proposed \textit{PPRNet} has a multi-scale structure, HUBs are embedded at different scales of the network to facilitate the utilization of the physics information to guide the training and inferencing of the network. It separately processes the global and local information of the feature maps with the aid of the Fourier intensity measurement and combines them with a channel attention method. Our ablation study has shown the importance of HUB in \textit{PPRNet}.
    
    \noindent\textbf{3.} Different from the traditional deep learning-based PR approaches that are trained and tested with simulation data, we construct an optical platform to evaluate the proposed \textit{PPRNet} and compare it with the state-of-the-art approaches. The results are thus more reliable to reflect the true performances in practical applications.  
    
    This paper is organized as follows. Section \ref{Sec:RelatedWorks} provides a comprehensive review of the traditional optimization-based algorithms and deep learning-based PR approaches. Section \ref{Sec:Approach} introduces the proposed \textit{PPRNet}. The simulation results and ablation studies are shown in Section \ref{Sec:Simulation}. The experiment results and comparisons with the existing methods are shown in Section \ref{Sec:Experiment}, followed by the conclusion in Section \ref{Sec:Conclusion}.
    
    \section{Related Works} \label{Sec:RelatedWorks}
    
    \subsection{Optimization-based PR Algorithms}
    In the early days, the traditional phase retrieval methods were based on the iterative alternating minimization (AM) framework \cite{Gerchberg1972APA, Fienup:82, Luke_2004}. The estimated image $\Tilde{\mathbf{x}} \in \mathbb{C}^{N\times N}$ is updated iteratively between the spatial and Fourier domain. Although the AM framework offers a portable solution, the AM-based algorithms are prone to stagnation and slow to converge (usually, more than $1000$ iterations are needed). Besides, they are sensitive to initialization.
    
    In recent years, the Writinger flow (WF) PR algorithm was developed with the advancement of modern optimization theories \cite{candes2015phase}. Different from the AM framework, WF solves the phase retrieval problem through gradient descent:
    \begin{equation}
    	\begin{aligned}
    		\Tilde{\mathbf{x}}^{k + 1} := \Tilde{\mathbf{x}}^{k} - \mu^{k + 1} \nabla f\left(\Tilde{\mathbf{x}}^{k}\right),
    	\end{aligned}
    \end{equation}
    where $\nabla f(\mathbf{x})$ represents the first-order gradient descent of the loss function (i.e. MSE loss), and $\mu^{k + 1}$ is the step size at current iteration. Empirically, $4$ to $8$ measurements are needed for a globally optimal solution \cite{candes2015code, candes2015phase}. Although WF provides a theoretical guarantee for convergence to the globally minimal solution, it often fails to converge to a satisfactory result if only one intensity measurement is given.
    
    \subsection{Deep Learning-based PR Methods}
    
    In recent years, the deep learning-based PR approaches have been widely studied since they provide non-iterative inferences compared with the time-consuming optimization-based algorithms. Most of these approaches can work with only one Fourier intensity measurement for each PR reconstruction without the need for any additional constraints on the measurement. This seemingly impossible task, in fact, has a theoretical basis \cite{hayes1982reconstruction}. It is known that if the Fourier intensity measurement is oversampled by two or more times in each dimension, the original complex-valued signal can be uniquely reconstructed, except for trivial ambiguities. Although such a reconstruction problem is highly non-convex (it is the reason why the traditional optimization methods fail to perform), it is particularly suitable to the deep learning-based methods due to their non-linear nature. Besides, they can also make use of the statistics acquired from a huge dataset to infer the solution. In general, the deep learning-based PR methods can be split into two categories depending on whether the underlying physics is adopted in the networks.
    
    For the first category, a feedforward network is used to estimate the target images directly from a Fourier intensity measurement \cite{Wu_cw5029, Wu_2021, uelwer2021phase, Uelwer_PhaseRetrieval, Nishizaki_2020, Wang_2020, Sinha17, Ye_SiSPRNet}. Specifically, \cite{Wu_cw5029} proposed a two-branch CNN to reconstruct the magnitude and phase part from an oversampled Fourier intensity measurement, and \cite{Wu_2021} applied this network to the 3D crystal PR problem. The approach shows reasonably good performance for simple images with brief details. The performance when dealing with complex images is unknown. \cite{uelwer2021phase} applied the conditional generative adversarial network to reconstruct the images. The network is extremely large since multiple multi-layer perceptrons (MLP) are used for all stages of the network. \cite{Nishizaki_2020} adopted the ResNet \cite{He_2016_CVPR} structure for Fourier phase retrieval tasks. Only a simple dataset (MNIST) was used for testing, and the performance was only barely satisfactory. The error in the details was still rather large. \cite{Sinha17} implemented a UNet structure \cite{ronneberger2015u} to reconstruct the phase-only images from Fresnel diffraction patterns. Its performance with Fraunhofer diffraction patterns (Fourier intensity) is unknown.  As for \cite{Wang_2020}, the authors adopted a multiple-resolution UNet structure and connected the hidden layers in the decoder to additional convolution layers to produce coarse outputs in an attempt to match the low-frequency components. Only the result of using $2$ measurements is shown in the paper, and the blurring effect is quite significant, as shown in the result. Recently, our team also developed a feedforward DNN structure to tackle the PR problem \cite{Ye_SiSPRNet}. It has an MLP front end for feature extraction and a residual attention-based reconstruction unit to generate the phase images. Although it outperformed most of the existing state-of-the-art methods, its performance when dealing with more complex images still had room for further improvement.  
    
    To improve the quality of the reconstructed images, the second category of deep learning-based PR methods implicitly or explicitly utilizes the underlying physics in the models \cite{DeepPhaseCut_2021, Zhang_21, I_l_2019, Morales_22, pmlr_metzler18a, hypernet_2022}. For instance, \cite{Morales_22} made use of the physics information to perform a learnable spectral initialization \cite{candes2015phase}. It is followed by a double branch UNet for reconstruction. The approach requires an additional masking scheme to impose constraints on the measurement. The reconstructed images are rather noisy, even for simple images. \cite{Uelwer_PhaseRetrieval} proposed to use MLP of different sizes in a cascaded network. The intensity measurement is applied to each MLP to assist the training and inferencing. The network size is very large due to the use of multiple MLPs. And the approach fails to reconstruct the details in the images. There are other iterative approaches similar to the traditional optimization-based methods. For instance, \cite{I_l_2019} suggested an iterative method with a 3-step structure: HIO initialization, iterative update between UNet and HIO, and final refinement by UNet.  \cite{pmlr_metzler18a} proposed to embed a pre-trained DnCNN \cite{zhang2017beyond} into a plug-and-play iterative algorithm for refining the estimated images at each iteration. These iterative physics-driven approaches are usually quite time-consuming. Besides, they cannot be end-to-end trained, which often affects the overall performance. 

\section{The Proposed Approach} \label{Sec:Approach}

\begin{figure*} [htb]
	\centering
	\includegraphics[width=\linewidth]{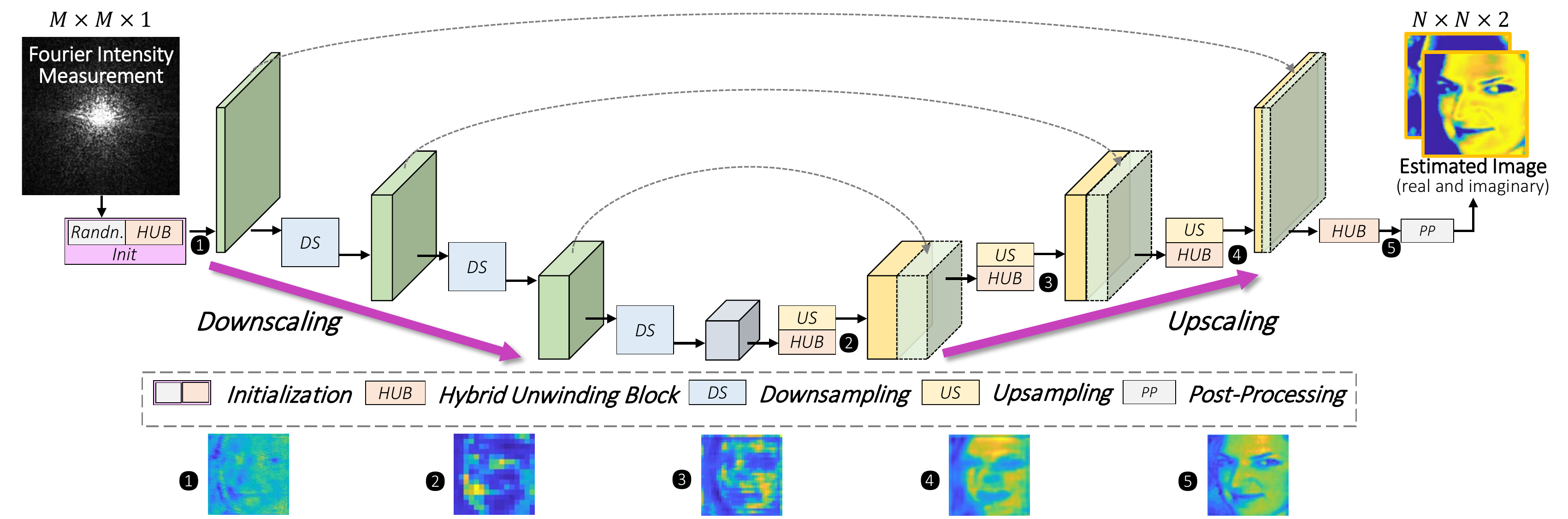}
	\caption{Architecture of the proposed Physics-Driven Phase Retrieval Network (\textit{PPRNet}). The most significant output feature maps after each HUB are shown at the bottom of the figure.}
	\label{fig:Network}
\end{figure*}

We present in this section the proposed Physics-Driven Phase Retrieval Network, \textit{PPRNet}. Fig. \ref{fig:Network} illustrates the overall architecture, which is essentially a multi-scale structure consisting of five parts: Initialization (Init), Hybrid Unwinding Block (HUB), Downsampling (DS), Upsampling (US), and Post-Processing (PP). The multi-scale structure has been generally applied to many deep learning-based image restoration tasks \cite{ronneberger2015u}. It effectively extracts the essential features of the input image while redundant and insignificant components, such as noises and outliers, are ignored. The essential features are then used to reconstruct the target image gradually in scale. When applying it to the Fourier PR problems, the input (Fourier intensity measurement) and required output (complex-valued spatial image), however, have a large domain discrepancy. It requires special add-on structures to accomplish the task. The details are explained below.

\subsection{The Overall Architecture} \label{sec:overallnetwork}

The proposed \textit{PPRNet} takes the Fourier intensity measurement $\bm{\mathcal{X}}\in\mathbb{R}^{M\times M\times1}$ as the input. It is fed to the Init Block to give the initial guess $\hat{\mathbf{x}}\ \in\ \mathbb{R}^{N\times N\times C}$ in the spatial domain, where $N<M/2$ and $C$ is the number of channels. In our experiment, $M,\ N$, and $C$ are set as $762$, $128$, and $64$, respectively. In the Init Block, a random image is first generated and refined by a HUB. As will be explained later, HUB is a learnable network structure and physics-informed. It is the core component of the proposed \textit{PPRNet}. Similar to \cite{ronneberger2015u}, the multi-scale structure consists of a contracting and an expanding path. In the contracting path, the size of the feature map is gradually reduced by several DS blocks to extract the essential context features. A DS block reduces the feature size by half through $3$ Convolution Blocks (ConvB). Each ConvB contains $1$ convolution layer with filter of size $3\times3$, $1$ Instance Norm layer, and a LeakyReLU activation function. The strides of the convoluional layers in the three Convolution Blocks are set to $2$, $1$ and $1$, respectively.  The resulting feature maps are then used in the expanding path to reconstruct the target complex-valued images. In the expanding path, the feature maps are enlarged gradually by the US Blocks to the original size using the nearest neighbor interpolation function. Most importantly, HUBs are introduced at different scales in the expanding path to carry out the reconstruction with the underlying physics. Finally, the PP Block reconstructs the image from the resulting feature maps by two convolutional layers. Similar to the traditional multi-scale structure, skip connections are introduced to allow the expanding path to use the fine-grained details learned in the contracting path to reconstruct the image. 

\subsection{Initialization} \label{sec:Init}
Initialization is a crucial step in traditional optimization-based algorithms as it provides an appropriate starting point in the spatial domain for the optimization process. Proper initialization can help the optimization algorithms avoid getting trapped in undesirable saddle points  \cite{candes2015phase}. Deep learning models, which are data-driven systems, are usually sensitive to the quality of the inputs. Improper initialization brings additional noise that harms the performance of the deep learning model. Instead of using conventional initialization approaches like \cite{Fienup:82, candes2015phase}, we propose to include the initialization into the learnable network structure. First, we randomly generate the spatial images $\Tilde{\mathbf{x}}_{init}\sim \mathcal{N}(0, 1)  \in \mathbb{R}^{N\times N}$. $\Tilde{\mathbf{x}}_{init}$ is then fed into a network that contains a $1 \times 1$ convolutional layer to convert the data into $C$ channels. They are then sent to a HUB to generate the initial guess for the subsequent multi-scale network. Since HUB is physics-informed and trained end-to-end with the other parts of the network, it gives a better initial estimation of the target image than the traditional approaches that only rely on the available physics information. An example is shown in Fig. \ref{fig:Network}. It can be seen that the shape of the object is roughly constructed by the Init block. It is left to the subsequent multi-scale structure to further enhance the image.

\subsection{Hybrid Unwinding Block (HUB)} \label{sec:HUB}

\begin{figure} [htb]
	\centering
	\includegraphics[width=\linewidth]{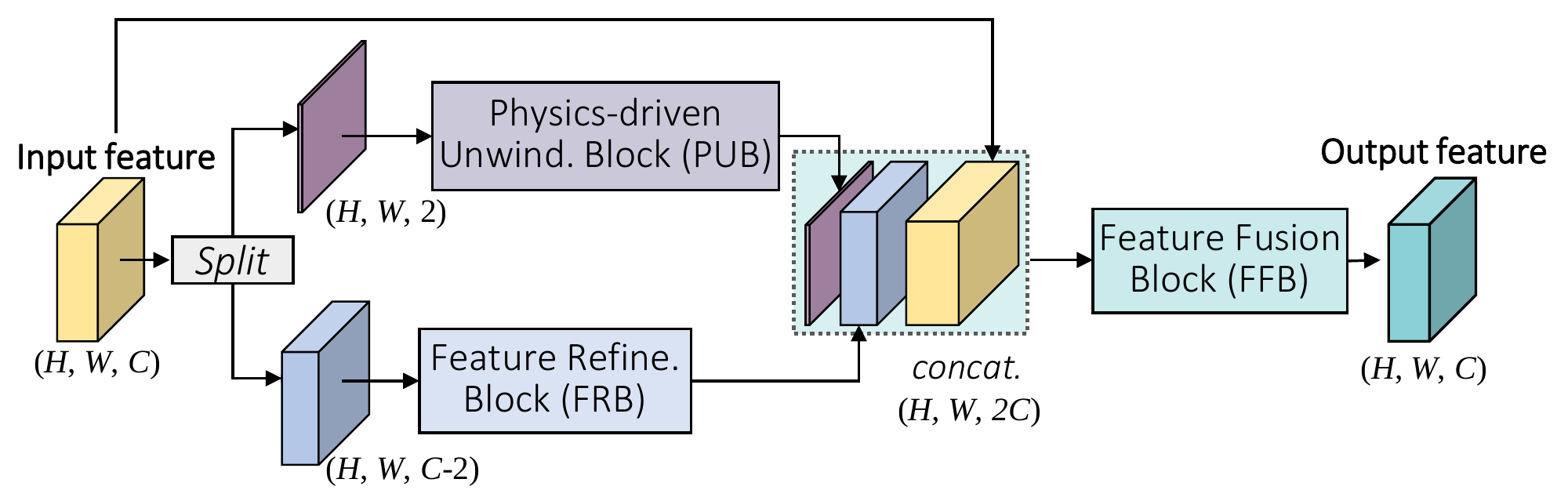}
	\caption{Structure of the Hybrid Unwinding Block (HUB).  }
	\label{fig:HUB}
\end{figure}

HUB is the crucial processing block of \textit{PPRNet}. It is used in the Init Block to refine the initial guess and also in the expanding path to guide the reconstruction process. Its structure is shown in Fig. \ref{fig:HUB}. HUB firstly splits the input feature maps $\Tilde{\mathbf{x}} \in \mathbb{R}^{H\times W\times C}$ into two branches for processing (where $H$ and $W$ are set as $N$ in our experiments). $\mathbf{u} \in \mathbb{R}^{H\times W\times 2}$ are the first two channels of the feature maps. They are fed to the Physics-driven Unwinding Block (PUB). And the rest channels $\mathbf{v} \in \mathbb{R}^{H\times W\times (C-2)}$ are processed by the Feature Refinement Block (FRB). We denote the outputs of PUB and FRB as $\mathbf{\Tilde{u}}$ and $\mathbf{\Tilde{v}}$, respectively. They are concatenated together with the input feature maps $\widetilde{\mathbf{x}}$ and sent to the Feature Fusion Block (FFB), which makes use of the channel attention method to extract the significant features for sending to the next stage. The detailed operations of these functional blocks are described below. 

\subsubsection{\textbf{Physics-driven Unwinding Block (PUB)}} \label{sec:PUB}
\begin{figure} [htb]
	\centering
	\includegraphics[width=\columnwidth]{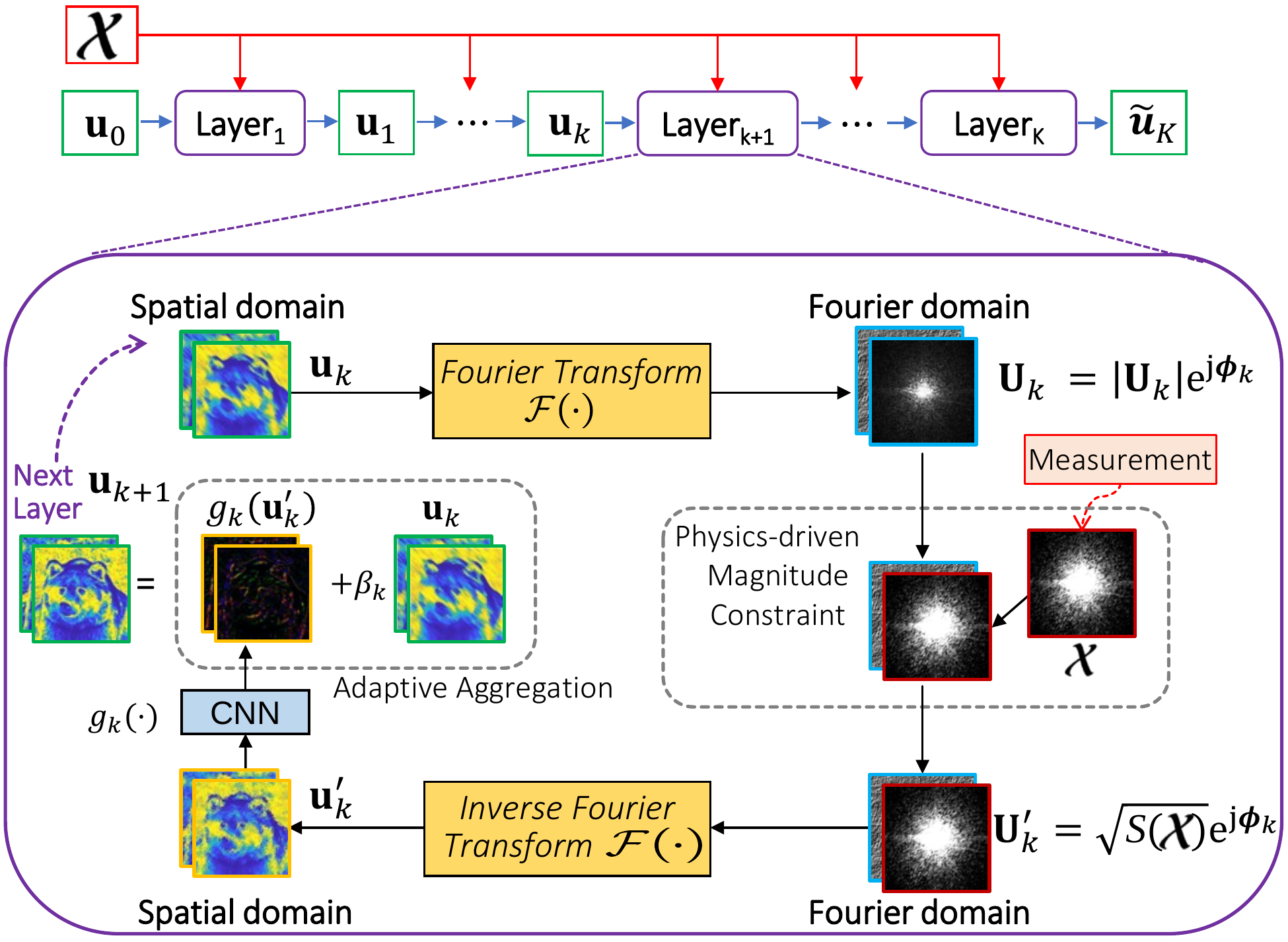}
	\caption{Structure of the Physics-driven Unwinding Block (PUB).}
	\label{fig:PUB}
\end{figure}

As shown in the traditional physics-driven PR methods, applying the prior domain knowledge to the optimization process can improve the estimation accuracy. A typical approach is to use the intensity measurement as prior information to constrain the optimization. It, however, often ends up with an iterative process since the constraint needs to be repeatedly applied to give the effect. To solve the problem, we propose PUB that unwinds the iterative process into a feedforward network operation. Fig. \ref{fig:PUB} shows the structure of PUB. It can be seen that a PUB contains $K$ unwinding layers cascaded in series. In our experiment, $K = 5$. Let us take the $\left(k+1\right)$th unwinding layer for illustration. $\mathbf{u}_k{\in\mathbb{R}}^{H\times W\times2}$, which forms the real and imaginary parts of the input feature map, are first transformed to the Fourier domain through: 
\begin{equation}
	\mathbf{U}_k=\left|\mathbf{U}_k\right|e^{j\phi_k}=\mathcal{F}\left(\mathbf{u}_k\right).
\end{equation}
We then update the magnitude with the intensity measurement to constrain the estimation. Note that HUB is applied to different scales in the expanding path. We need to convert the intensity measurement to the respective scale for applying to PUB. The updated $\mathbf{U}_k'$ is then mapped back to the spatial domain through the inverse Fourier transform and produces an updated image $\mathbf{u}_k'$. The whole operation can be expressed as:
\begin{equation}
	\mathbf{u}_k'=\mathcal{F}^{-1}\left(\sqrt{S\left(\bm{\mathcal{X}}\right)}e^{j\phi_k}\right),
\end{equation}
where $S\left(\bm{\mathcal{X}}\right)$ refers to the filtering and folding of $\bm{\mathcal{X}}$ for converting it to the required scale without aliasing. It forms the magnitude constraint that brings domain-specific prior information to the reconstruction process during the training and inference of the network. It informs the network of the reconstruction target at every scale of the expanding path. $\mathbf{u}_k'$ is then fed to a shallow CNN structure $g_k\ (.)$ (empirically, we stacked $8$ convolutional layers) for learning the missing phase information. Instead of generating the refined image directly, $g_k$ is trained to give the residue of the desired output from the input feature map. It reduces the training difficulty. The operation can be expressed as follows:
\begin{equation}
	\mathbf{u}_{k+1}=g_k\left(\mathbf{u}_k'\right)+\beta_k\mathbf{u}_k,
\end{equation}
where $\beta_k$ is a learnable parameter. The resulting image $\mathbf{u}_{k+1}$ then acts as the input to the next layer of PUB. Both $g_k\left(\cdot\right)$ and $\beta_k$ can be trained adaptively to control the amount of magnitude constraint to be included in $\mathbf{u}_{k+1}$. Together with FFB (which will be described later), they provide the flexibility to the network to determine how much the physics information should be utilized in the image reconstruction process through end-to-end training. 

\subsubsection{\textbf{Feature Refinement Block (FRB)}}
Images contain both global and local features. The PUB illustrated above aims to constrain the image reconstruction through the prior Fourier intensity measurement. It is well-known that the Fourier transform can only give the global frequency information of an image. The magnitude constraint alone cannot effectively inform the local features in the image. Therefore, in parallel with PUB, we propose a Feature Refinement Block (FRB) to enrich the detailed structures corresponding to the local features. FRB consists of three ConvBs. The shallow FRB structure can learn detailed representations from the input feature maps, for example, high-frequency details like complex shapes and edges. With the global information from PUB and local textures from FRB, the concatenated feature maps (\textit{concat.} in Fig. \ref{fig:HUB}) can provide comprehensive representations for the reconstruction process.

\subsubsection{\textbf{Feature Fusion Block (FFB)}}
\begin{figure} [htb]
	\centering
	\includegraphics[width=\linewidth]{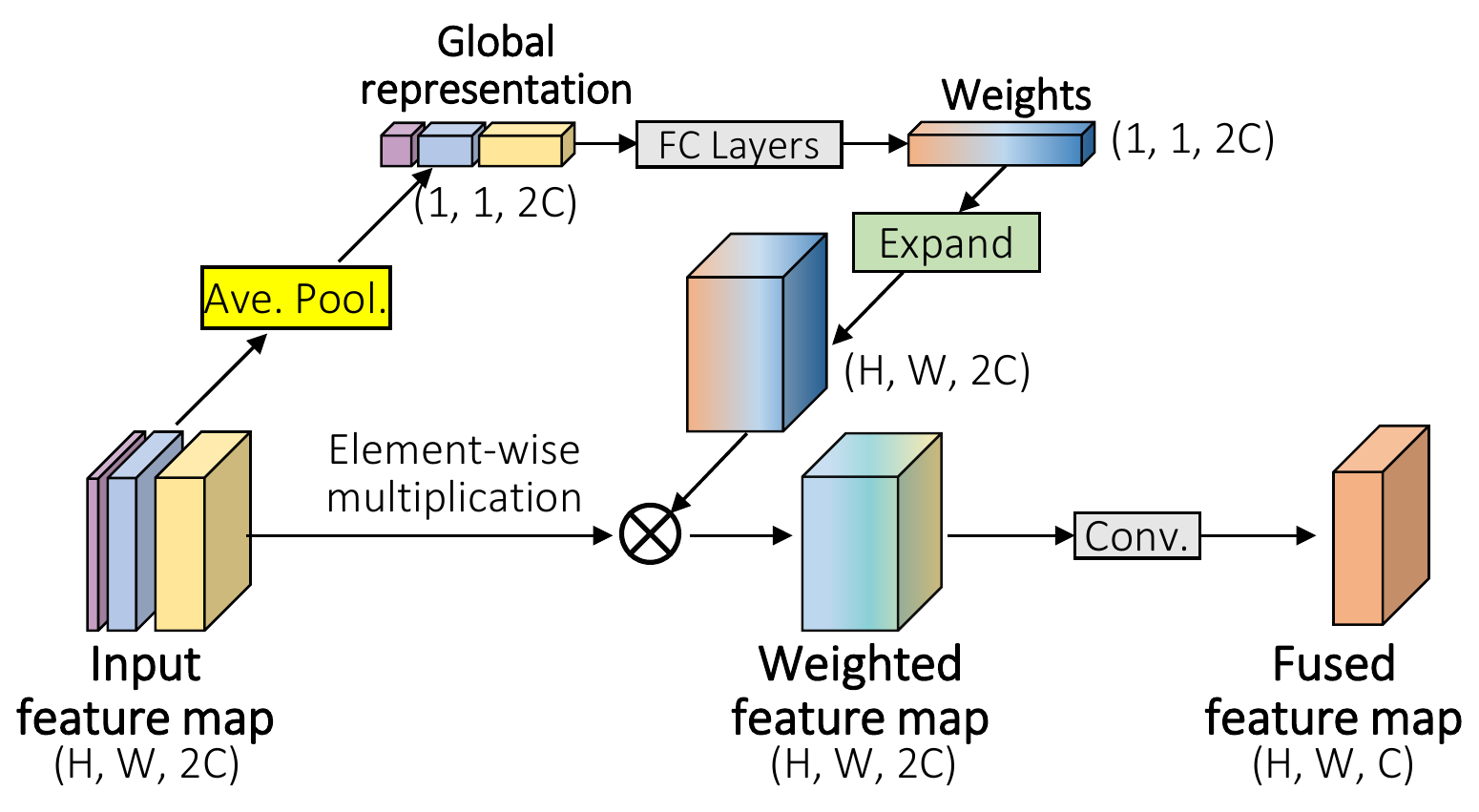}
	\caption{Structure of the Feature Fusion Block (FFB). }
	\label{fig:FFB}
\end{figure}

As shown in Fig. \ref{fig:HUB}, the output feature maps of PUB and FRB are concatenated with the input feature maps to form the resulting feature maps with size ($H,W, 2C$). The simple concatenation cannot effectively utilize the representations of the feature maps. Obviously, these $2C$ channels of feature maps have different importance to the reconstruction. We propose to use a Feature Fusion Block (FFB), which is essentially a channel attention network  \cite{Hu_2018_CVPR}, to adaptively re-weight these channels based on their contents. They are then combined to form the output of HUB. More specifically, the structure of FFB is shown in Fig. \ref{fig:FFB}. The input feature maps are first sent to an average pooling layer (Ave. Pool in Fig. \ref{fig:FFB}) to find the global representation of different channels. It uses a single value extracted by the average pooling layer to represent the global information of each channel. In other words, a total of $2C$ values are used to represent the input feature maps with $2C$ channels. Then, two fully-connected layers (FC layers in Fig. \ref{fig:FFB}) work together to investigate the channel-wise dependency and produce $2C$ weights to indicate the importance of these channels. The key features have large weights since they are essential for the reconstruction. The weight of each channel is expanded $H\times W$ times to have the same spatial dimension as the input feature map, i.e., the Expand block in Fig. \ref{fig:FFB}. They are then element-wise multiplied with the input feature maps to give different levels of attention. The re-weighted feature maps are fused together by a ConvB to have the size of $H\times W\times C$. They become the output of HUB and also the reconstructed images for a particular scale. Note that the input of FFB contains the feature maps generated by PUB (with the physics information),  FRB, and those from the contracting path. The feature maps of PUB do not necessarily play a key role in the final output, although they are physics informed. The physics information is adopted only when it helps the final reconstruction, determined by FFB, as informed by the end-to-end training. Fig. \ref{fig:Network} and Fig. \ref{fig:Vis_HUB} show some examples of the reconstructed images at different scales. Note that only the most significant feature map of each scale is shown. It is seen that the quality of the reconstructed image gradually improves as the scale increases. 

\subsection{Loss Function}
We regard the phase retrieval task as a supervised learning problem. There is a corresponding complex-valued image $\mathbf{x}$ as ground-truth reference given a Fourier intensity measurement $\bm{\mathcal{X}}$. Thus, we can measure the difference between ground-truth image $\mathbf{x}$ and the image $\mathbf{\hat{x}}$ estimated by \textit{PPRNet}. The loss function $\mathcal{L}$ we used contains two parts: pixel-wise loss, $\mathcal{L}_{pixel}$, minimizes the pixel-wise difference between the estimation and ground-truth images; and total variation (TV) loss, $\mathcal{L}_{TV}$, uses a smoothness prior to regularize the estimated image while maintaining its edges and textures. The total loss function $\mathcal{L}$ is formulated as:

\begin{equation}
	\mathcal{L}\left(\mathbf{\hat{x}}, \mathbf{x}\right) = \mathcal{L}_{pixel}(\mathbf{\hat{x}}, \mathbf{x}) + \gamma \mathcal{L}_{TV}(\mathbf{\hat{x}}) , 
	\label{Eq:lossfunc}
\end{equation}
where $\gamma \in \mathbb{R}$ denotes the coefficient to balance different loss terms. $\mathcal{L}_{pixel}$ is defined as the $\ell_1$-norm distance between the estimation and ground-truth images. It promotes the fidelity of the images estimated by \textit{PPRNet} while preserving the details of the original image. $\mathcal{L}_{pixel}$ is expressed as:
\begin{equation}
	\mathcal{L}_{pixel}(\mathbf{\hat{x}}, \mathbf{x}) = \frac{1}{2N^2}\sum_{i=1}^{N} \sum_{j=1}^{N} \sum_{k=1}^{2} \left(\left| x^{Re}_{i,j,k}-\hat{x}^{Re}_{i,j,k} \right| + \left| x^{Im}_{i,j,k}-\hat{x}^{Im}_{i,j,k} \right| \right).
\end{equation}
The terms $\left(\cdot\right)^{Re}$ and $\left(\cdot\right)^{Im}$ represent the real and imaginary parts of the image, respectively. 

TV norm sums all the gradients along the horizontal and vertical directions. Using the TV regularization encourages the spatial smoothness in the estimated image such that it can assist in noise reduction and increase the consistency of the reconstructed image. $\mathcal{L}_{TV}$ is defined as:
\begin{equation}
	\mathcal{L}_{TV}(\mathbf{\hat{x}}) = \frac{1}{2N^2}\sum_{i=1}^{N} \sum_{j=1}^{N} \sum_{k=1}^{2}\left(\left(\hat{x}_{i, j, k}-\hat{x}_{i+1, j, k}\right)^{2}+\left(\hat{x}_{i, j, k}-\hat{x}_{i, j+1, k}\right)^{2}\right).
\end{equation}

    \section{Simulation and Ablation Studies} \label{Sec:Simulation}
    
    \subsection{Defocus-Based Fourier Phase Retrieval System} \label{sec:PRsystem}
\begin{figure} [ht]
    	\centering
    	\includegraphics[width=\linewidth]{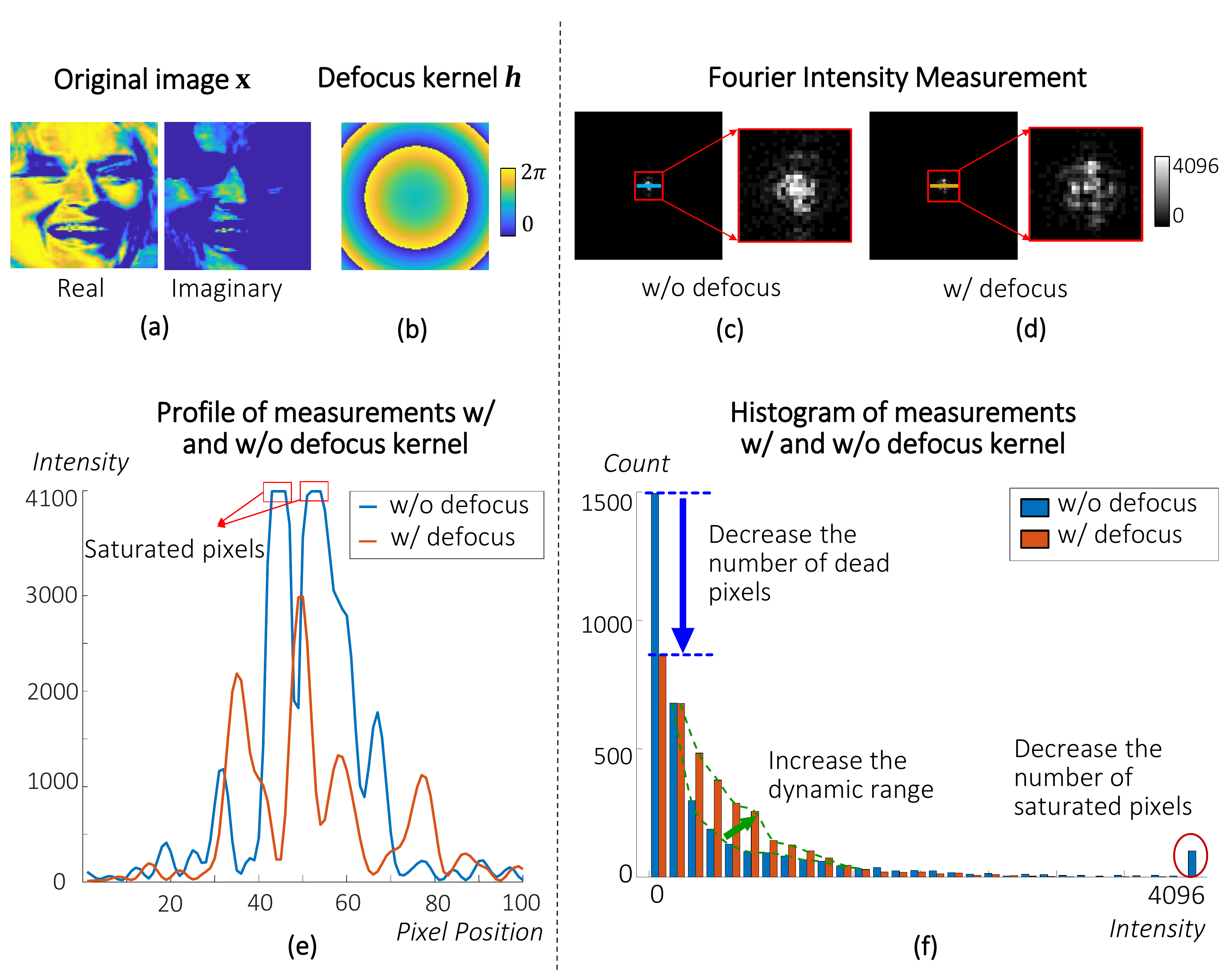}
    	\caption{(a) A sample complex-valued image (real and imaginary parts). (b) Visualization of the defocus kernel (phase part). The Fourier intensity measurement (collected by a scientific camera) of the sample image (c) without the defocus kernel and (d) with the defocus kernel. (e) Profile of the colored lines in (c) and (d) (with and without the defocus kernel). (f) Histogram of intensity measurements in (c) and (d). Zoom in for a better view.}
    	\label{fig:compdefocus}
    \end{figure}

    A Fourier phase retrieval system reconstructs the image $ \mathbf{x}\in \mathbb{C}^{N\times N}$ from its Fourier intensity measurement $\bm{\mathcal{X}}\in \mathbb{R}^{M\times M}$. However, as mentioned in Section \ref{Sec:RelatedWorks}, the saturation problem often happens when directly capturing Fourier intensity images using standard imaging devices. It is due to the large dynamic range of Fourier intensity data. An example is given in Fig. \ref{fig:compdefocus}. Fig. \ref{fig:compdefocus}(a) shows a typical structured image (complex-valued). Its Fourier intensity measurement obtained using a $12$-bit dynamic range camera is shown in Fig. \ref{fig:compdefocus}(c), with the central region magnified for better visualization. The profile along the blue line is shown in Fig. \ref{fig:compdefocus}(e) (blue line). It has a flat top due to the saturation problem. We also show the histogram of the intensity image in Fig. \ref{fig:compdefocus}(f). It can be seen that there are many pixels having the maximum value due to the saturation problem. Besides, we can also find many pixels having zero values. Most of them come from the high-frequency parts of the measurement. They have very small values as compared with the low-frequency data. They are thus quantized to zero, resulting in the so-called dead pixels. Consequently, the image contains many errors in both the low-frequency and high-frequency regions. 
    
\begin{figure} [htb]
    	\centering
    	\includegraphics[width=\linewidth]{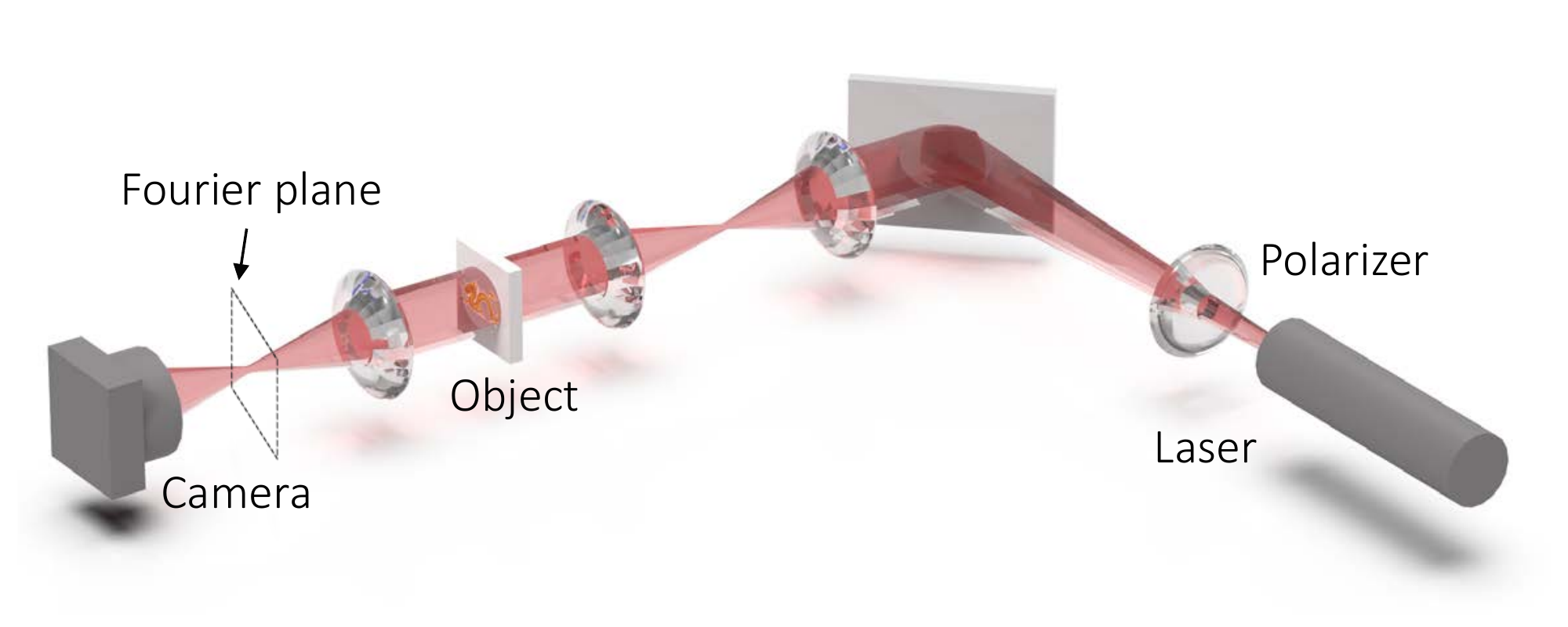}
    	\caption{The optical path of the defocus-based PR system.}
    	\label{fig:opticalpath}
    \end{figure}

    Researchers have suggested a few solutions to the problem. One of them is by using the defocusing method \cite{Ye_SiSPRNet}. Specifically, we can reduce the dynamic range of the intensity measurement by convolving it with a defocus kernel $\mathbf{H}$. The convolution operation can be easily implemented by moving the camera beyond the Fourier plane. Fig. \ref{fig:opticalpath} shows a typical optical path of a defocus-based PR system. In the figure, the object of interest is illuminated by a coherent light generated by a laser beam. The camera is placed beyond the focal plane such that a defocused Fourier intensity is captured. Mathematically, we denote the original image and its Fourier transform as $x(p, q)$ and $X(u,\ v)$, respectively. Then, the optical field $X_L(u', v')$ on the defocus plane can be formulated as: 
    \begingroup
    \allowdisplaybreaks
    \begin{align*}
    	X_L(u', v') & =  \frac{ e^{jkL} }{ jkL } e^{\frac{-jk}{2L} (u^{'2} +v^{'2})} \iint X(u, v) H(u - \frac{u'}{\lambda L}, v - \frac{v'}{\lambda L}) du dv\\
    	& =  C \iint \left[ \iint X(u, v) e^{j2\pi (x'u+y'v)} du dv \right] \\ 
    	&  \left[ \iint H(f_1, f_2) e^{j2\pi (x'f_1+y'f_2)} df_1 df_2 \right]e^{\frac{-jk}{L} (u'x' + v'y')} dx' dy'\\
    	& = C\  \mathcal{F}_{\lambda L}\left(\mathbf{x}\circ \mathbf{h}\right)(u', v')\numberthis \label{Eq:defocus}
    \end{align*}
    \endgroup
    where $\lambda$ and $k = \frac{2\pi}{\lambda}$ denote the wavelength and wave number, respectively. The symbol $L$ represents the distance between the camera and the Fourier plane. The term $\mathbf{H}$ is the Fourier transform of the defocus kernel $\mathbf{h}$. $C$ is a constant and $h(p, q)$ is defined as $e^{\frac{j\pi}{\lambda L}(p^{2}+q^{2})} $ \cite{goodman2017introduction}. As shown in \eqref{Eq:defocus}, the defocusing is equivalent to the element-wise multiplication of the original image and defocus kernel $\mathbf{h}$ in the spatial domain with the scaling factor $\lambda L$. An example of the defocus kernel is shown in Fig. \ref{fig:compdefocus}(c). Thus in our experiment, we directly implement $\mathbf{h}$ together with the testing images on an SLM. More details will be provided in Section \ref{Sec:Experiment}. 
    
\begin{figure}[htb]
    	
    	\begin{subfigure}{\linewidth}	
    		\centering
    		\includegraphics[width=0.7\linewidth]{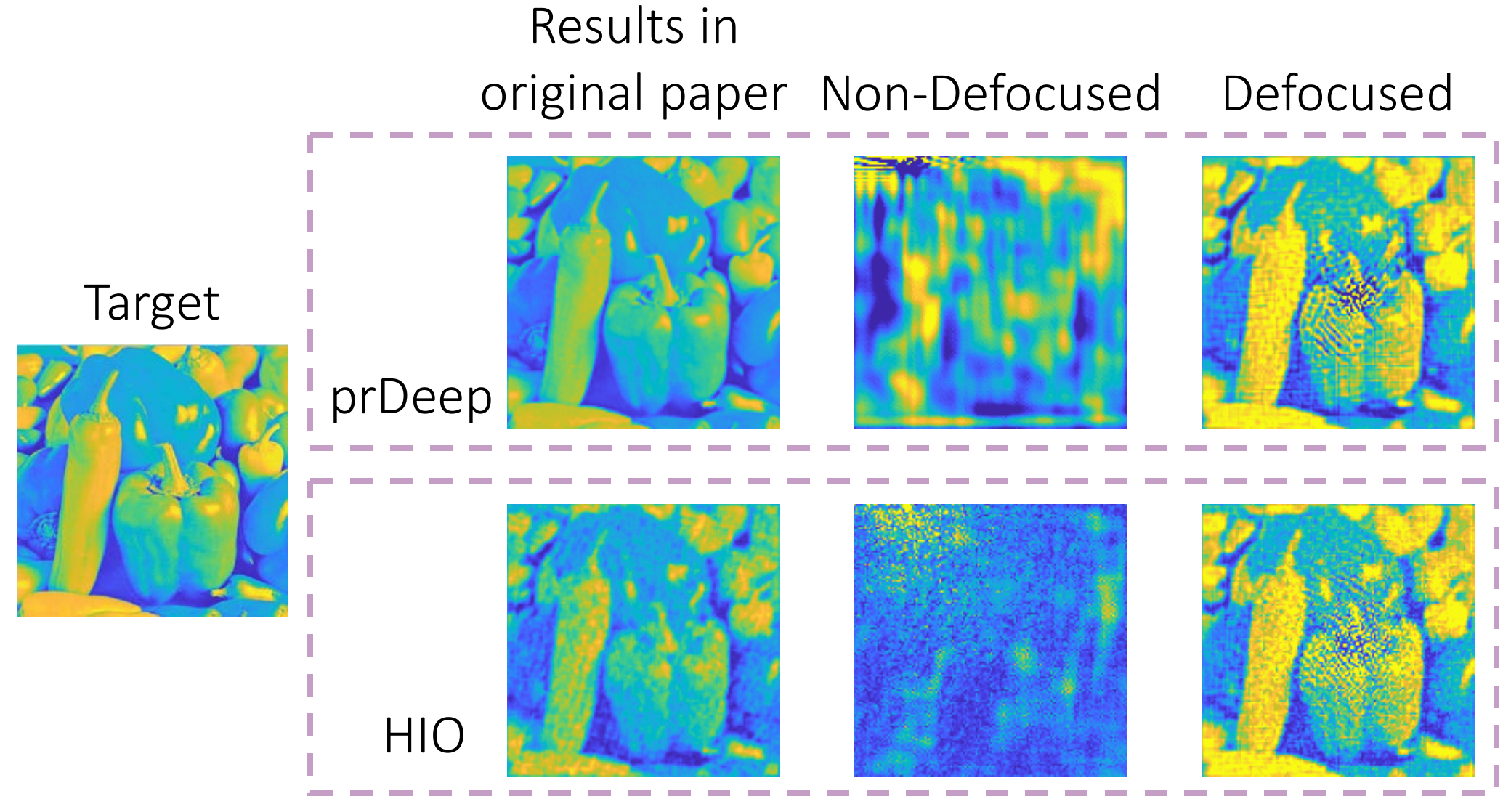}
    		\vspace*{-2mm}
    		\caption{}
    	\end{subfigure}
    	
    	\begin{subfigure}{\linewidth}	
    		\centering
    		\renewcommand{\arraystretch}{1.2}
    		\begin{adjustbox}{width=\linewidth,center}
    			\begin{tabular}{ccc|cc|cc}
    				\hline \multicolumn{1}{c}{ \multirow{2}{*}{ Methods } }  & \multicolumn{2}{c|}{ Noisy (original paper) } & \multicolumn{2}{c|}{ Non-Defocused} & \multicolumn{2}{c}{ Defocused } \\
    				& PSNR $\uparrow$ & SSIM $\uparrow$ & PSNR $\uparrow$ & SSIM $\uparrow$ & PSNR $\uparrow$ & SSIM $\uparrow$  \\
    				\hline
    				HIO \cite{Fienup:82} & $20.61$ & $ 0.40$ & $-5.65$ & $0.03$ & $11.31$ & $0.34$  \\
    				PrDeep \cite{pmlr_metzler18a} & $27.43$ & $0.68$ & $3.37$ & $0.14$ & $12.08$  & $0.21$\\
    				\hline
    			\end{tabular}
    		\end{adjustbox}
    		\caption{}
    	\end{subfigure}
    	\vspace{-2mm}
    	\caption{(a) Qualitative and (b) quantitative evaluation results of HIO \cite{Fienup:82} and PrDeep \cite{pmlr_metzler18a} with, and without using the defocused kernel. The quantitative results are the average of $6$ natural images used in \cite{pmlr_metzler18a}.}    
    	\label{fig:prdeepHIO}
    \end{figure}

    With the defocusing method, the saturation problem and dead pixels are greatly reduced, as shown in Fig. \ref{fig:compdefocus}(e) and (f). Although the defocused intensity measurement is not the exact intensity measurement of the image, PR methods using the defocused intensity measurements usually perform much better than using the original measurements with saturated and dead pixels. Some examples are given in Fig. \ref{fig:prdeepHIO}. In the figure, we first show the performance of the HIO \cite{Fienup:82} and PrDeep \cite{pmlr_metzler18a} methods, which represent the optimization-based and deep learning-based PR methods. They were implemented following the setting in their original papers (i.e., without saturation and dead pixels). The results are similar to those reported in \cite{pmlr_metzler18a}. Then, we introduced the saturation problem and dead pixels to the intensity measurements by capping the image data as $12$-bit integers. It can be seen that the performances of these methods drop substantially when the saturation problem and dead pixels in the measurements are taken into account. It shows the results of the traditional approaches using only the simulation data without considering these problems are unreliable for practical applications. Finally, we used the defocusing method as mentioned above to mitigate the problem. As shown in Fig. \ref{fig:prdeepHIO}, both approaches can significantly benefit from using the defocusing method. An improvement of up to $15$ dB can be achieved. For this reason, in this work, all comparing methods are trained and tested with the defocused intensity measurements.
    
    \subsection{Simulation} \label{sec:simulation}
    
    \subsubsection{Datasets} \label{sec:simulationsetup} To verify the effectiveness of the proposed \textit{PPRNet}, we conducted comprehensive simulations with complex-valued images. To prepare these images for training and testing, we first collected images from two publicly available datasets: Real-world Affective Faces (RAF) dataset \cite{li2017reliable} and Fashion-MNIST dataset \cite{xiao2017/online}. The images were converted to grayscale and resized to $128\times128$ pixels. We then constructed two datasets by combining these images differently. The first dataset has linearly correlated magnitude and phase parts. It was constructed by using the images in the Fashion-MNIST dataset. For each image obtained from the dataset $\mathbf{x}_{\text{\textbf{raw}}}$, it was scaled to $\left[0,\ 1\right]$ and used as the magnitude part $\mathbf{x}_{mag}$ of the final image $\mathbf{x}$. For the phase part, we applied an exponential function to obtain it, i.e.,         $\mathbf{x}_{phase}=exp\left(2\pi i\mathbf{x}_{\text{\textbf{raw}}}\right)$. Finally, we combined the magnitude and phase parts by $\mathbf{x}=\mathbf{x}_{mag}\circ x_{phase}$, where $\circ$ denotes the element-wise multiplication. We used the first $25000$ images of the Fashion-MNIST’s training dataset to create our training set and used the first $1000$ images of its testing dataset to create our testing set. The second dataset we constructed contains images with uncorrelated magnitude and phase parts. We applied the same approach mentioned above to re-scale the data but used images from two different datasets, namely, RAF and Fashion-MNIST, to generate the magnitude and phase parts, respectively. There were $12771$ images in the training set and $1000$ images in the testing set.
    
    To simulate the defocusing effect as discussed in Section \ref{sec:PRsystem}, we multiplied all images by a defocus kernel, which was generated by a built-in function of Holoeye SLM control software corresponding to the defocusing distance of $30mm$. The same defocus kernel was used in the experiments on the optical platform, as will be discussed later. Then, $2$-D FFT was performed on these images, and the magnitudes of the resulting images were extracted to become the intensity measurements. The resolution of the measurements is $762\times762$ pixels. To simulate the saturation and quantization errors, we capped the measurements with the maximum limit $4095$ ($12$-bit) and converted the numbers to the integer format.  
    
    \subsubsection{Training Details} We trained our network with the Adam optimizer \cite{kingma2017adam}. The learning rate was set to $10^{-4}$, and the batch size was set to $24$. The network was trained for $160$ epochs on the PyTorch deep learning platform on a PC with two NVIDIA RTX3090 GPUs. The weighting factor $\gamma$ in the loss function \eqref{Eq:lossfunc} was empirically set to $0.1$. The number of layers in PUB was set as $5$ for the speed-accuracy trade-off. More discussion on our choices can be found in the ablation studies.
    
    \subsubsection{Metrics and Evaluation} We used the Peak Signal to Noise Ratio (PSNR, higher the better), Structural Similarity (SSIM, higher the better), and Mean Absolute Error (MAE, lower the better) as the performance criteria to measure the discrepancy between the reconstructed images $\hat{\mathbf{x}}$ and ground truth images $\mathbf{x}$. They are widely used in the image restoration field to measure the estimation quality. The phase parts of all images are shifted by $\pi$ to ensure that no negative values are used for the computation of PSNR. The average PSNR, SSIM, and MAE for all $1000$ testing images of both datasets are used for comparison.
    
    \subsubsection{Compared Methods}To evaluate the effectiveness of the proposed \textit{PPRNet}, we compare its performance with a few state-of-the-art methods, namely, HIO \cite{Fienup:82}, which represents the traditional optimization-based method; 
    PRCGAN \cite{uelwer2021phase}; HIO-UNet \cite{I_l_2019}; LenlessNet \cite{Sinha17}; NNPhase \cite{Wu_cw5029}; and MCNN \cite{Wang_2020}, which represent the deep learning-based approaches. Note that some of these approaches are designed under much trivial input and output requirements as compared with those in this paper. For instance, NNPhase and PRCGAN assume the input intensity measurement is very small (only $64\times 64$  and $28\times 28$ pixels, respectively, as compared to $762\times 762$ in the simulation environment). And PRCGAN assumes the target image only has real-valued data (as compared to having both real and imaginary data in the simulation environment). We need to modify these networks slightly to let them adapt to the simulation environment. Specifically, for PRCGAN and NNPhase, we first convert these networks to accept $128\times 128$ pixels input. Then, we put a pre-processing block that contains two convolutional layers with a $5\times 5$ kernel and strides $3$ and $4$, respectively, in front of the original network to convert the dimension of the input data from $762\times 762$ to $128\times 128$ pixels. Besides, we modified the output of PRCGAN to have $2$ channels for magnitude-phase representation, and HIO-UNet and LenlessNet to have $2$ channels for real-imaginary representation. These modifications allow these networks to perform in the simulation environment. 
    
    \subsubsection{Simulation Results}
\begin{figure*}[htb]
    	
    	\begin{subfigure}{\textwidth}	
    		\centering
    		\begin{adjustbox}{width=\textwidth,center}
    			\begin{tabular}{*{15}{c@{\extracolsep{0.05em}}} }
    				\includegraphics[width=0.08\textwidth,valign=t]{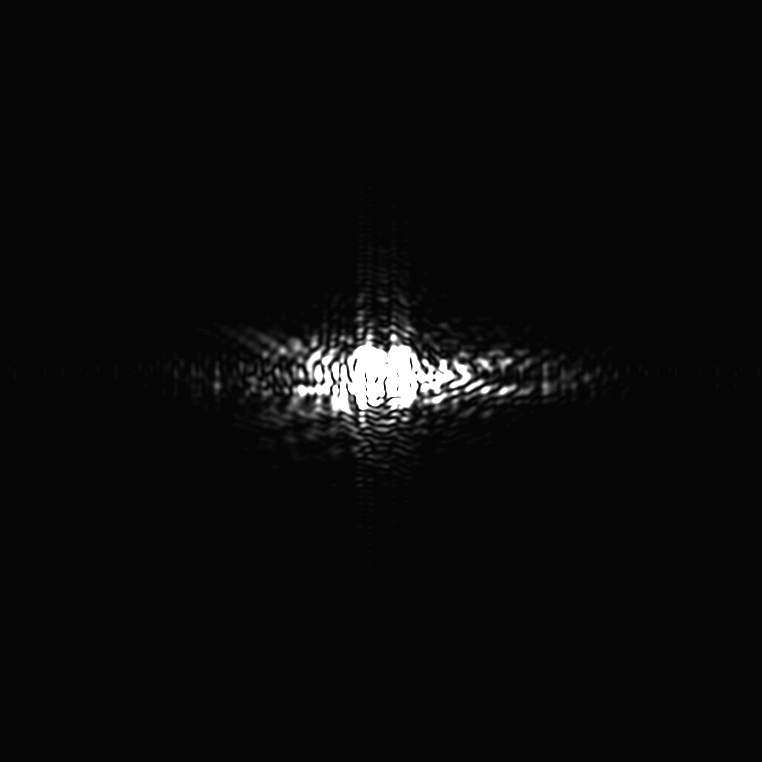}
    				&\includegraphics[height=0.08\textwidth,valign=t]{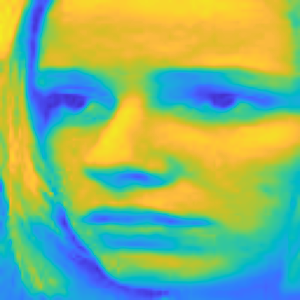}
    				&\includegraphics[width=0.08\textwidth,valign=t]{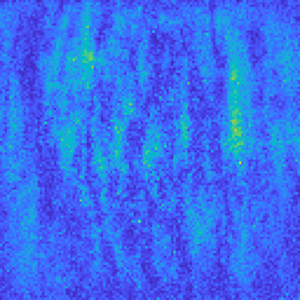}
    				&\includegraphics[width=0.08\textwidth,valign=t]{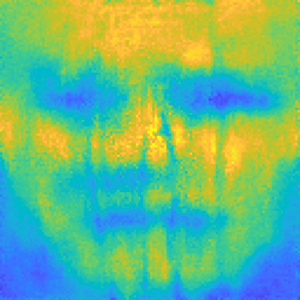}
    				&\includegraphics[width=0.08\textwidth,valign=t]{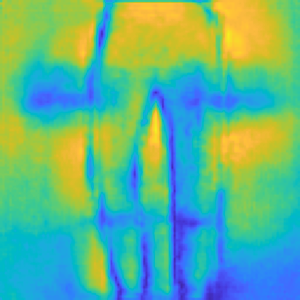}
    				&\includegraphics[width=0.08\textwidth,valign=t]{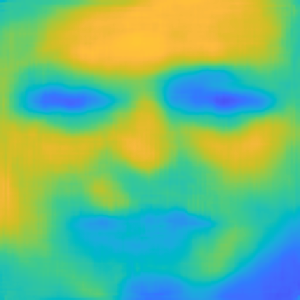}
    				&\includegraphics[width=0.08\textwidth,valign=t]{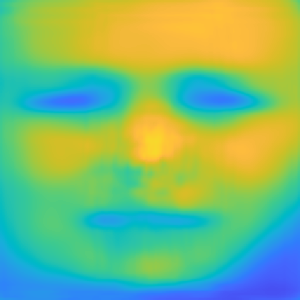}
    				&\includegraphics[width=0.08\textwidth,valign=t]{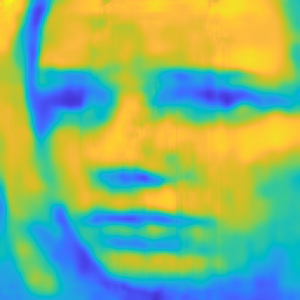}~
    				&\includegraphics[height=0.08\textwidth,valign=t]{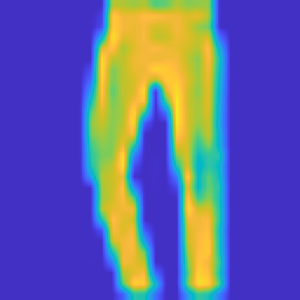}
    				&\includegraphics[width=0.08\textwidth,valign=t]{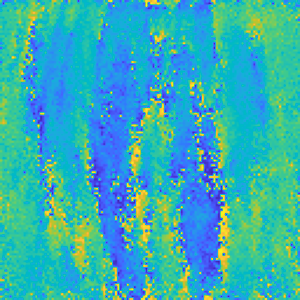}
    				&\includegraphics[width=0.08\textwidth,valign=t]{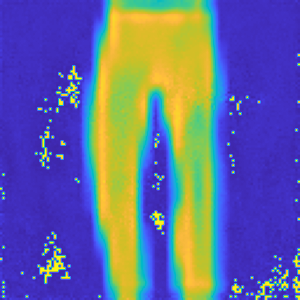}
    				&\includegraphics[width=0.08\textwidth,valign=t]{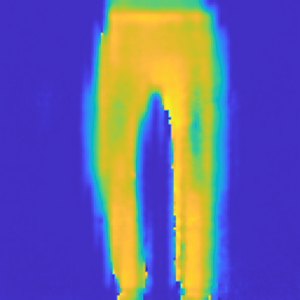}
    				&\includegraphics[width=0.08\textwidth,valign=t]{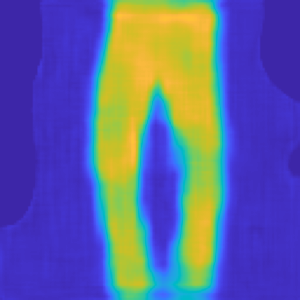}
    				&\includegraphics[width=0.08\textwidth,valign=t]{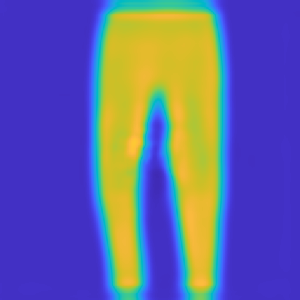}
    				&\includegraphics[width=0.08\textwidth,valign=t]{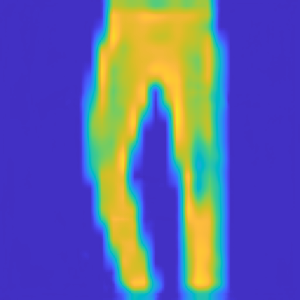}
    				
    				\\\addlinespace[0.2em]
    				
    				\includegraphics[width=0.08\textwidth,valign=t]{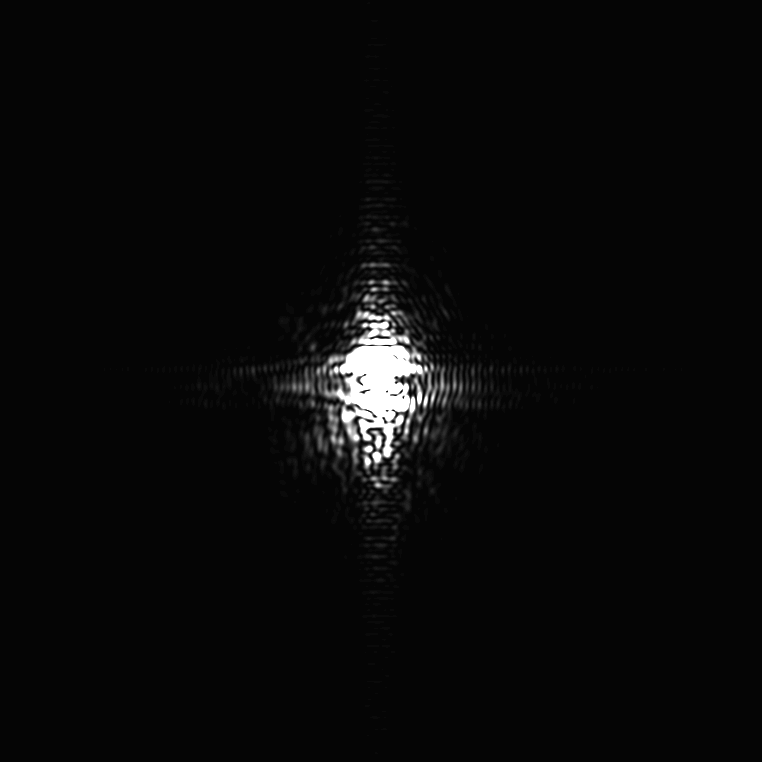}
    				&\includegraphics[height=0.094\textwidth,width=0.084\textwidth,valign=t]{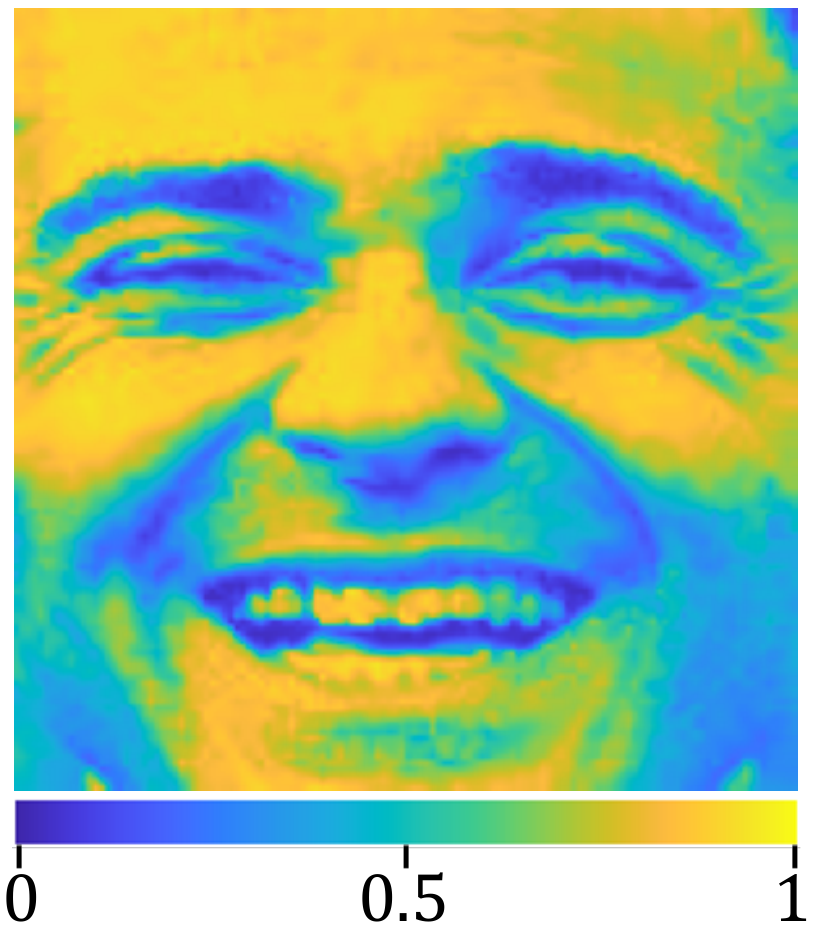}
    				&\includegraphics[width=0.08\textwidth,valign=t]{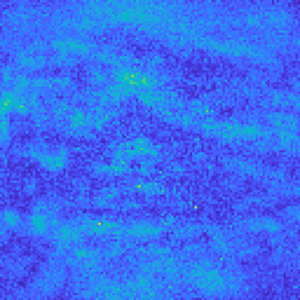}
    				&\includegraphics[width=0.08\textwidth,valign=t]{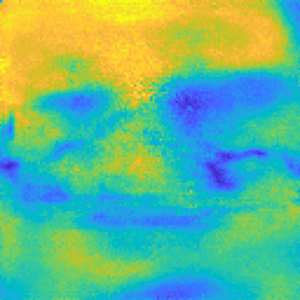}
    				&\includegraphics[width=0.08\textwidth,valign=t]{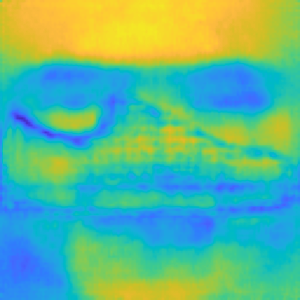}
    				&\includegraphics[width=0.08\textwidth,valign=t]{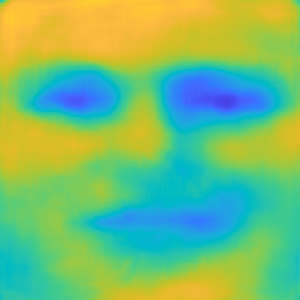}
    				&\includegraphics[width=0.08\textwidth,valign=t]{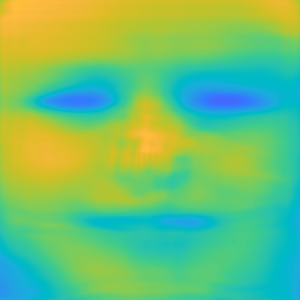}
    				&\includegraphics[width=0.08\textwidth,valign=t]{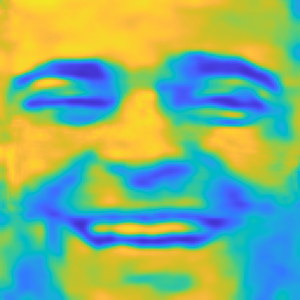}~
    				&\includegraphics[height=0.094\textwidth,width=0.084\textwidth,valign=t]{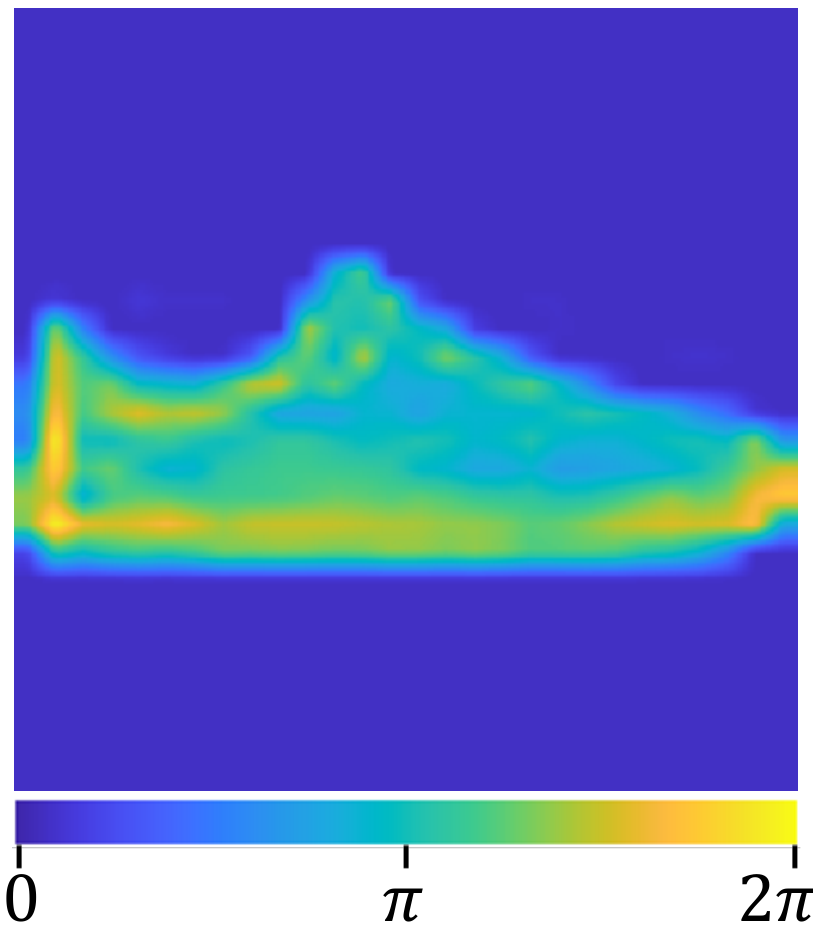}
    				&\includegraphics[width=0.08\textwidth,valign=t]{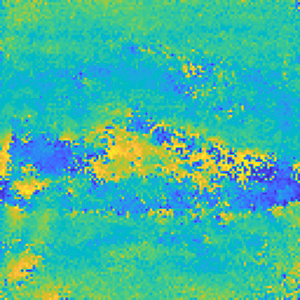}
    				&\includegraphics[width=0.08\textwidth,valign=t]{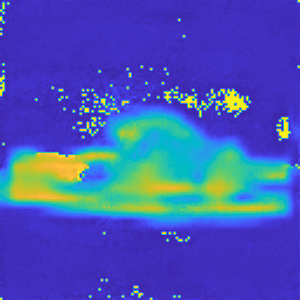}
    				&\includegraphics[width=0.08\textwidth,valign=t]{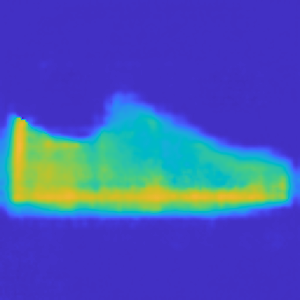}
    				&\includegraphics[width=0.08\textwidth,valign=t]{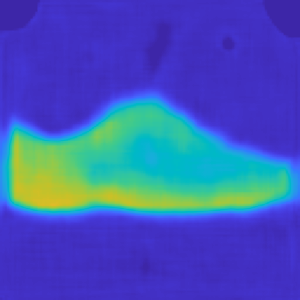}
    				&\includegraphics[width=0.08\textwidth,valign=t]{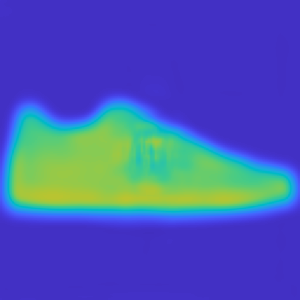}
    				&\includegraphics[width=0.08\textwidth,valign=t]{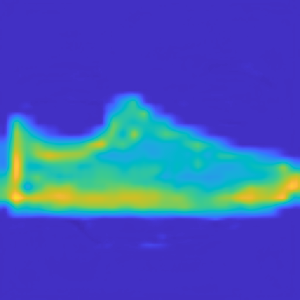}

    			\end{tabular}
    		\end{adjustbox}
    		\vspace*{-1.5mm}
    		\caption{}
    	\end{subfigure}

    	\begin{subfigure}{\textwidth}	
    		\centering
    		\begin{adjustbox}{width=\textwidth,center}
    			\begin{tabular}{*{15}{c@{\extracolsep{0.05em}}} }
    				
    				\includegraphics[width=0.08\textwidth,valign=t]{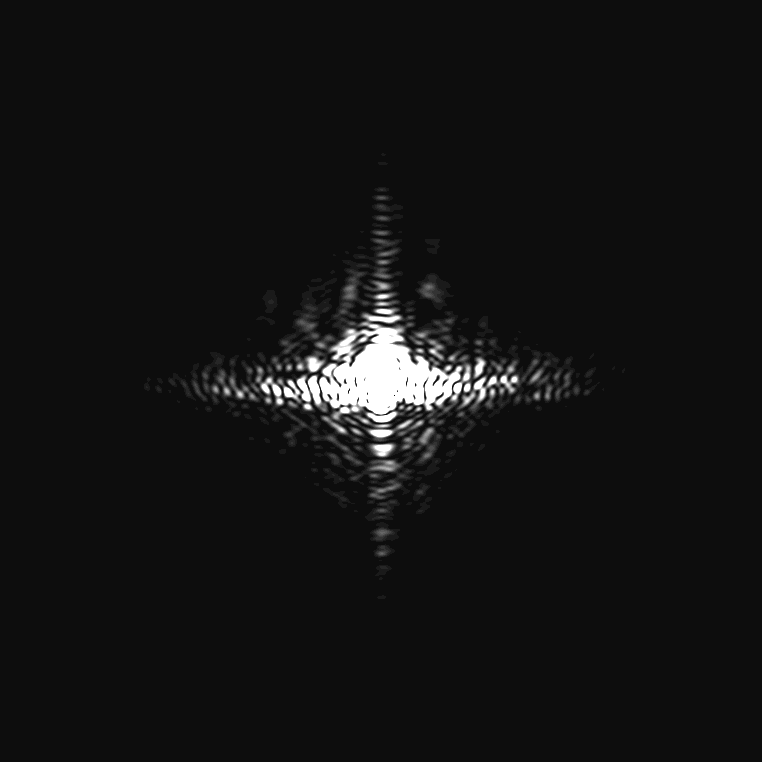}
    				&\includegraphics[height=0.08\textwidth, valign=t]{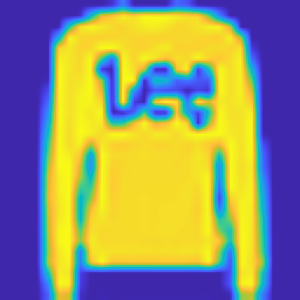}
    				&\includegraphics[width=0.08\textwidth,valign=t]{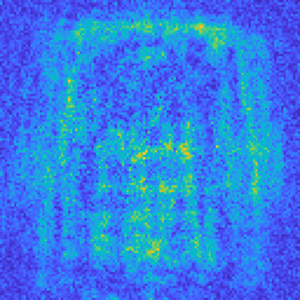}
    				&\includegraphics[width=0.08\textwidth,valign=t]{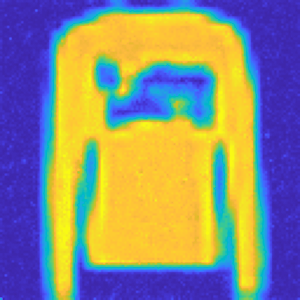}
    				&\includegraphics[width=0.08\textwidth,valign=t]{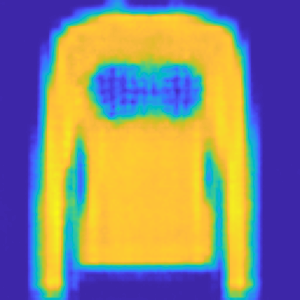}
    				&\includegraphics[width=0.08\textwidth,valign=t]{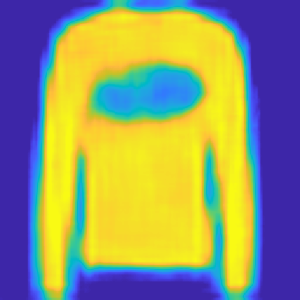}
    				&\includegraphics[width=0.08\textwidth,valign=t]{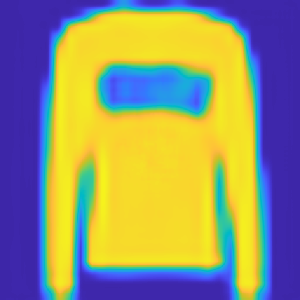}
    				&\includegraphics[width=0.08\textwidth,valign=t]{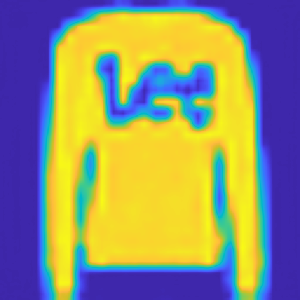}~
    				&\includegraphics[height=0.08\textwidth,valign=t]{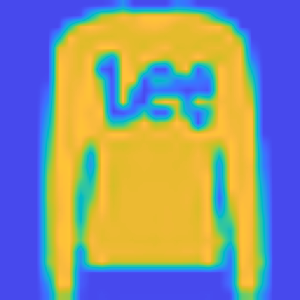}
    				&\includegraphics[width=0.08\textwidth,valign=t]{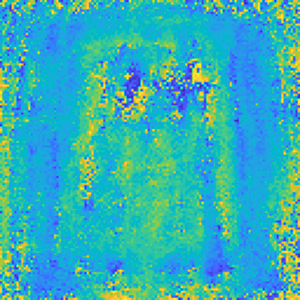}
    				&\includegraphics[width=0.08\textwidth,valign=t]{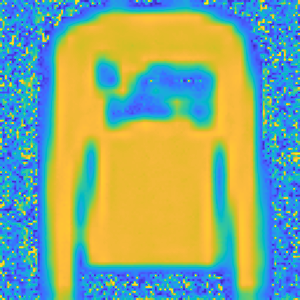}
    				&\includegraphics[width=0.08\textwidth,valign=t]{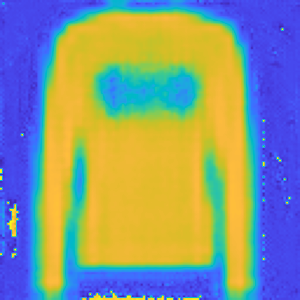}
    				&\includegraphics[width=0.08\textwidth,valign=t]{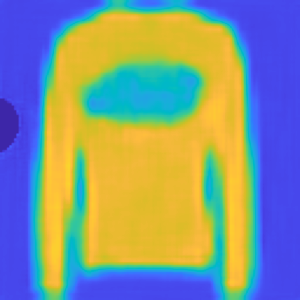}
    				&\includegraphics[width=0.08\textwidth,valign=t]{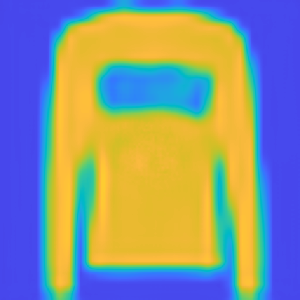}
    				&\includegraphics[width=0.08\textwidth,valign=t]{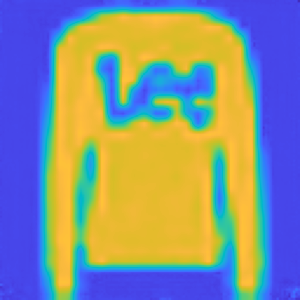}
    				\\\addlinespace[0.2em]
    				
    				\includegraphics[width=0.08\textwidth,valign=t]{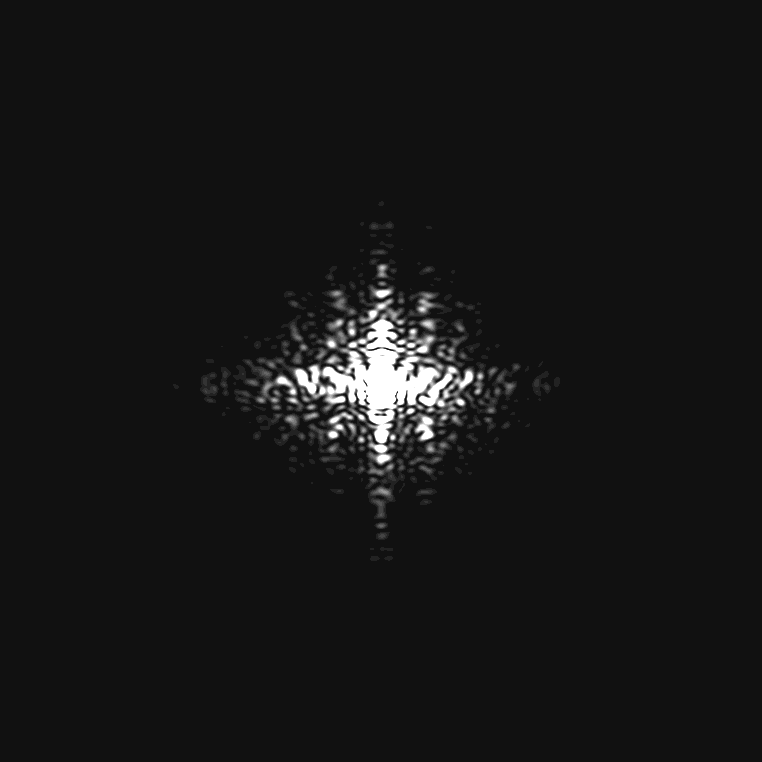}
    				&\includegraphics[height=0.094\textwidth,width=0.084\textwidth,valign=t]{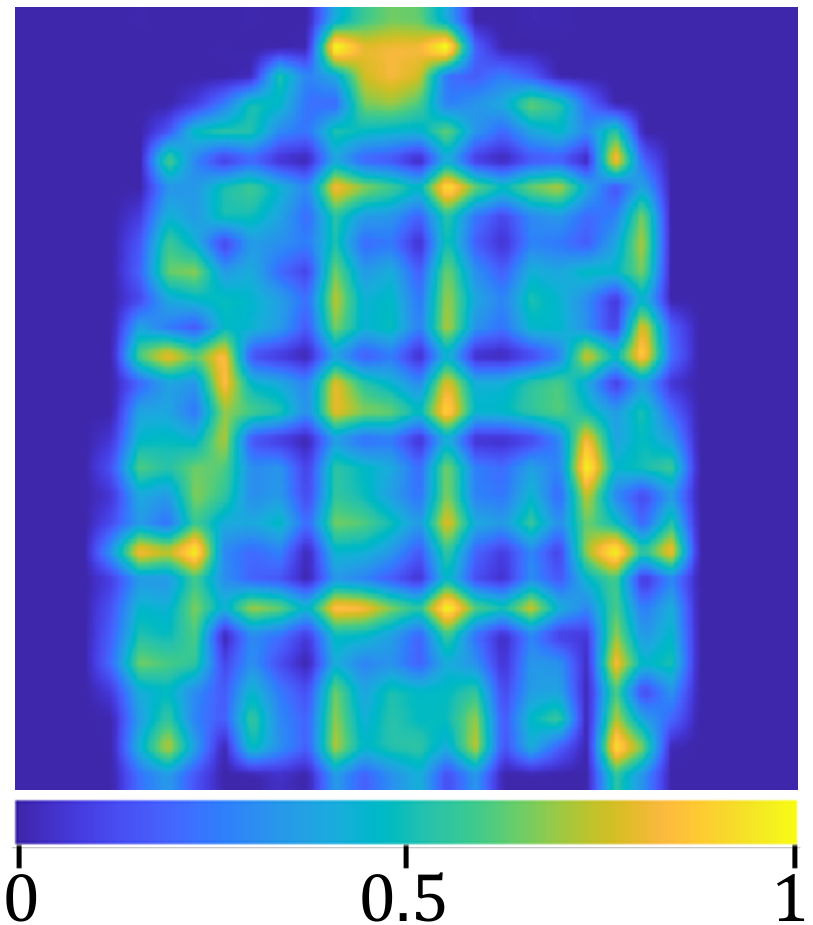}
    				&\includegraphics[width=0.08\textwidth,valign=t]{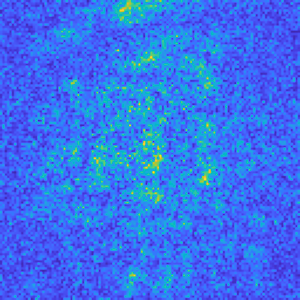}
    				&\includegraphics[width=0.08\textwidth,valign=t]{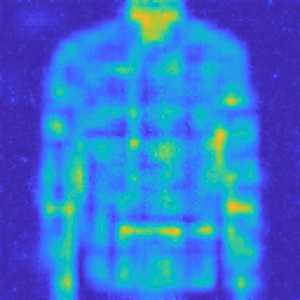}
    				&\includegraphics[width=0.08\textwidth,valign=t]{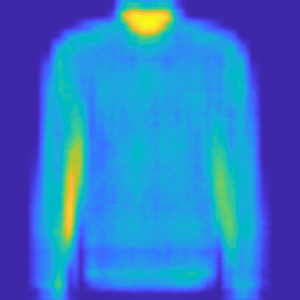}
    				&\includegraphics[width=0.08\textwidth,valign=t]{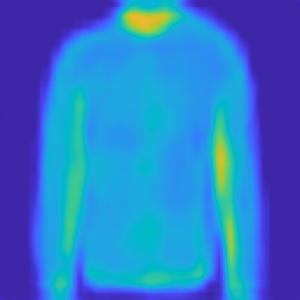}
    				&\includegraphics[width=0.08\textwidth,valign=t]{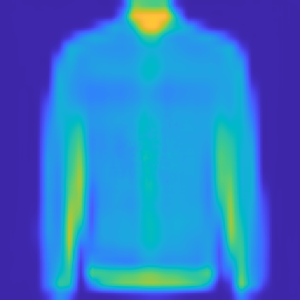}
    				&\includegraphics[width=0.08\textwidth,valign=t]{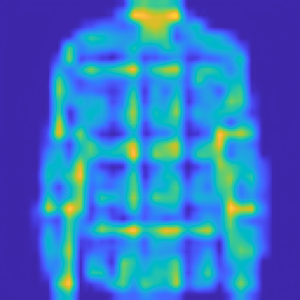}~
    				&\includegraphics[height=0.094\textwidth,width=0.084\textwidth,valign=t]{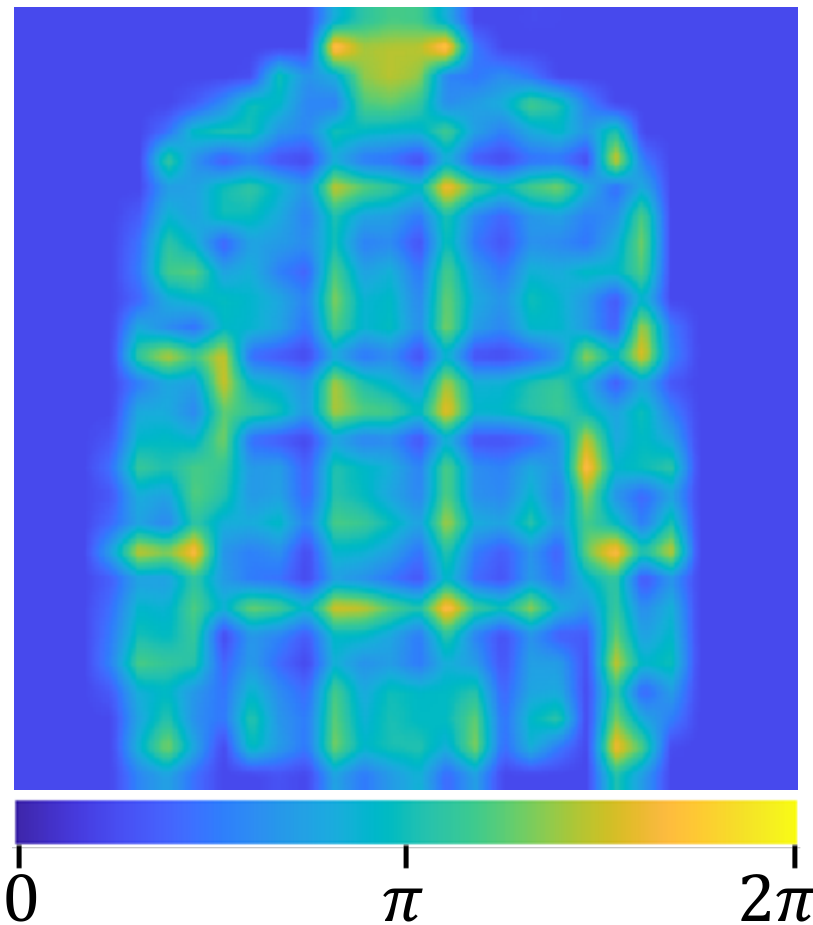}
    				&\includegraphics[width=0.08\textwidth,valign=t]{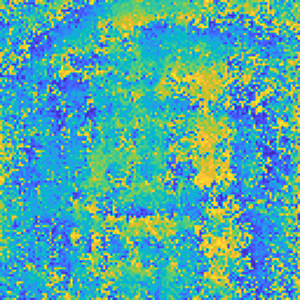}
    				&\includegraphics[width=0.08\textwidth,valign=t]{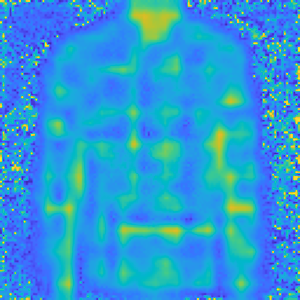}
    				&\includegraphics[width=0.08\textwidth,valign=t]{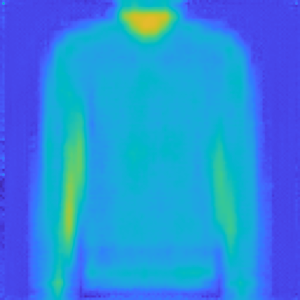}
    				&\includegraphics[width=0.08\textwidth,valign=t]{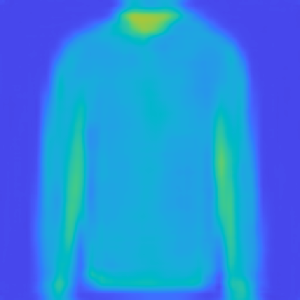}
    				&\includegraphics[width=0.08\textwidth,valign=t]{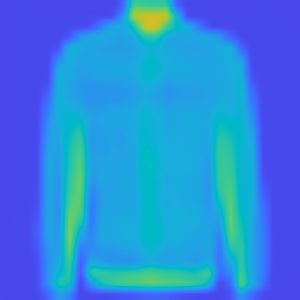}
    				&\includegraphics[width=0.08\textwidth,valign=t]{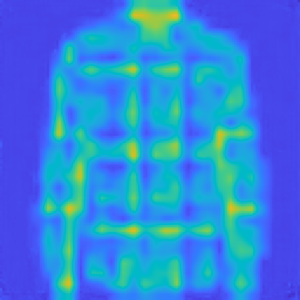}
    				\\\addlinespace[0.4em]
    				
    				{\scriptsize \makecell[c]{Fourier\\measurement}} & {\scriptsize \makecell[c]{Ground-truth \\ (mag.)}} & {\scriptsize \makecell[c]{HIO \\ (mag.)}} & {\scriptsize \makecell[c]{HIO-UNet \\ (mag.)}} &{\scriptsize \makecell[c]{PRCGAN \\ (mag.)}} & {\scriptsize \makecell[c]{NNPhase \\ (mag.)}} & {\scriptsize \makecell[c]{MCNN \\ (mag.)}} & {\scriptsize \makecell[c]{\textbf{\textit{PPRNet} }  \\ (mag.)}} & {\scriptsize \makecell[c]{Ground-truth \\ (phase)}}  & {\scriptsize \makecell[c]{HIO \\ (phase)}} & {\scriptsize \makecell[c]{HIO-UNet \\ (phase)}} & {\scriptsize \makecell[c]{PRCGAN \\ (phase)}} & {\scriptsize \makecell[c]{NNPhase \\ (phase)}} & {\scriptsize \makecell[c]{MCNN \\ (phase)}} & {\scriptsize \makecell[c]{\textbf{\textit{PPRNet}  } \\ (phase)}}
    				\\
    			\end{tabular}
    		\end{adjustbox}
    		\vspace*{-1mm}
    		\caption{}
    	\end{subfigure}
    	
    	\begin{subfigure}{\textwidth}	
    		\centering
    		\renewcommand{\arraystretch}{1.2}
    		\begin{adjustbox}{width=\textwidth,center}
    			\begin{tabular}{*{7}{c}|*{6}{c}}
    					\hline \multicolumn{1}{c}{ \multirow{3}{*}{ Methods } }  & \multicolumn{6}{c|}{ Uncorrelated } & \multicolumn{6}{c}{ Linear Correlated }  \\
    					& PSNR $\uparrow$ & PSNR $\uparrow$ & SSIM $\uparrow$ & SSIM $\uparrow$ & MAE $\downarrow$ &  MAE $\downarrow$  & PSNR $\uparrow$ & PSNR $\uparrow$ & SSIM $\uparrow$ & SSIM $\uparrow$ & MAE $\downarrow$ & MAE $\downarrow$  \\
    					& Mag. & Phase & Mag. & Phase & Mag. &  Phase   & Mag. & Phase  & Mag. & Phase  & Mag. & Phase \\ 
    					\hline
    					\multicolumn{13}{c}{Traditional Methods}\\ 
    					\hline HIO \cite{Gerchberg1972APA} & $-20.437$ & $7.729$ & $0.0002$ & $0.075$ & $9.329$ &  $1.760$  & $7.709$ & $8.317$ & $0.329$ & $0.043$ & $0.298$ & $1.898$ \\
    					
    					\hline \multicolumn{13}{c}{Learning-based Methods}\\ 
    					\hline 
    					LenlessNet \cite{Sinha17} & $17.394$ & $20.365$ & $0.412$ & $0.748$ & $0.106$ &  $0.359$  & $17.216$ & $17.807$ & $0.624$ & $0.432$ & $0.079$ & $0.594$  \\
    					PRCGAN \cite{uelwer2021phase} & $15.525$ & $20.163$ & $0.342$ & $0.725$ & $0.130$ &  $0.343$  & $21.452$ & $23.502$ & $0.719$ & $0.738$ & $0.047$ & $0.256$ \\
    					NNPhase \cite{Wu_cw5029} & $17.191$ & $18.876$ & $0.460$ & $0.613$ & $0.108$ &  $0.458$  & $21.697$ & $24.193$ & $0.690$ & $0.794$ & $0.049$ & $0.229$ \\
    					MCNN \cite{Wang_2020} & $17.880$ & $18.641$ & $0.483$ & $0.692$ & $0.098$ &  $0.424$  & $22.454$ & $24.986$ & $0.774$ & $0.819$ & $0.042$ & $0.201$ \\
    					HIO-UNet \cite{I_l_2019} & $15.597$ & $14.285$ & $0.294$ & $0.538$ & $0.132$ & $0.612$   & $22.464$ & $12.830$ & $0.541$  & $0.410$ & $0.054$ &   $0.853$\\
    					\rowcolor{pink!50} \textbf{\textit{PPRNet} (Ours)} & $\mathbf{23.698}$ & $\mathbf{31.698}$ & $\mathbf{0.694}$ & $\mathbf{0.933}$ & $\mathbf{0.053}$ &  $\mathbf{0.113}$   & $\mathbf{32.991}$ & $\mathbf{34.178}$ & $\mathbf{0.920}$ & $\mathbf{0.925}$ & $\mathbf{0.017}$ & $\mathbf{0.101}$   \\
    					\hline
    				\end{tabular}
    			\end{adjustbox}
    			\caption{}
    		\end{subfigure}
    		\vspace{-0.2cm}
    		\caption{Qualitative simulation results of different phase retrieval methods on two datasets with images having (a) uncorrelated and (b) linearly correlated magnitude and phase components. The pixel values of the Fourier intensity measurements (first column) range from $0$ to $4095$. The magnitude part of each target image and their corresponding colormaps are shown in the second column. The third to sixth columns denote the reconstructed magnitude parts through different methods. The seventh column presents the phase part of each target image and their corresponding colormaps. The reconstructed phase parts via different approaches are provided from the eighth to last columns. Zoom in for better view. (c) Quantitative comparisons (average PSNR/SSIM/MAE) on the two datasets. Best performances are denoted in \textbf{bold} font.}    
    		\label{fig:comp_magphase}
    	\end{figure*}

    The qualitative and quantitative comparison results are shown in Fig. \ref{fig:comp_magphase}. To save space, we do not include the qualitative results of LenslessNet in Fig. \ref{fig:comp_magphase} due to its inferior performance. All approaches give better performance for the dataset with linearly correlated magnitude and phase components, and the proposed \textit{PPRNet} outperforms all compared approaches. As shown in Fig. \ref{fig:comp_magphase}(b), only \textit{PPRNet} can reconstruct the details in the images. For instance, it is clearly seen that only \textit{PPRNet} can successfully recover the ”Lee” characters and plaid on the shirt. The images given by other approaches have poor quality. On the other hand, the dataset with uncorrelated magnitude and phase components is challenging for all competing methods, as shown in Fig. \ref{fig:comp_magphase}(a). It is particularly the case for PRCGAN and HIO-UNet since they originally were designed to output only real-valued output. The distortion in the magnitude images, as shown in Fig. \ref{fig:comp_magphase}(a), is quite obvious. For approaches like NNPhase and MCNN, the output images are rather blurry and dissimilar from the target images. These approaches also cannot reconstruct the details in the phase images. It is worth noting that the distortion of the HIO algorithm is quite severe. In general, the reconstruction of complex-valued objects is considerably more difficult than for real-valued, non-negative objects for the HIO algorithm \cite{Fienup:87}. It is particularly the case for our dataset where the images have uncorrelated magnitude and phase such that the object supports are difficult to define. Compared with the above approaches, the proposed \textit{PPRNet} gives the best performance. The reconstructed magnitude and phase images closely follow the target images, particularly the phase images. Although we can occasionally find some artifacts in the magnitude images, they are not serious. 
    
    The quantitative comparison is shown in Fig. \ref{fig:comp_magphase}(c). The proposed \textit{PPRNet} achieves much better performance than all compared methods evaluated by different metrics. For the magnitude part, the proposed \textit{PPRNet} achieves average PSNR and SSIM gains of at least $5.818dB$ and $0.146$, respectively. For the phase part, the increases in PSNR and SSIM gains can reach $9.192dB$ and $0.106$, respectively. The above simulation results show that the proposed \textit{PPRNet} outperforms the state-of-the-art methods. We will show in Section \ref{Sec:Experiment} that the same conclusion can be drawn when testing these methods in a practical environment.
    
    \subsection{Ablation Analysis}
    
    For selecting the hyperparameters and better understanding the roles of different components of the proposed \textit{PPRNet}, a series of ablation analyses were conducted. In these analyses, the RAF dataset was used to generate the required training and testing images. To allow the analyses to reflect the performance when testing with the optical setup (which will be described in Section \ref{Sec:Experiment}), we set the target images to have a constant magnitude but varied phase components (i.e., they are all phase-only images). So, when evaluating the network, only the phases of the reconstructed images are considered.
    
    \subsubsection{Effect of Multi-Scale Network}
    As introduced in Section \ref{sec:overallnetwork}, we proposed to use a multi-scale structure that includes a contracting and expanding path. Multiple scales can help improve the representation power of the feature maps and thus better exploit the image information. Besides, having more scales lets us include more HUBs into the expanding path. It is equivalent to having more iterations in the iterative deep learning PR methods that will generally give a better result. To investigate the effects of the multi-scale features, we conducted a set of quantitative comparisons. The results are shown in TABLE \ref{table:ablationscale}. It can be seen that the performance becomes better with the growth of the scale from one to three. It verifies our expectations mentioned above. However, the network deteriorates when the scale is larger than three since the resolutions of the lower scale feature maps are too small to represent the image appropriately. For instance, when the scale is four, the resolution of the lowest-scale feature map is just $8 \times 8 (256/32)$. It is too small to represent the image. Besides, the number of parameters and operations will also increase when there are more scales. Thus, we chose to have $3$ scales in the proposed \textit{PPRNet}.
    
    \begin{table}[ht]
    	\centering
    	\caption{Comparison among using different number of scales for the proposed \textit{PPRNet} on the RAF dataset (simulation). The best performances are marked in \textbf{bold} font.}
    	\renewcommand{\arraystretch}{1.2}
    	\setlength\tabcolsep{7.8pt}
    	\begin{tabular}{cccccc}
    		\hline
    		\# Scales &
    		PSNR $\uparrow$ &
    		SSIM $\uparrow$ &
    		MAE $\downarrow$ &
    		\begin{tabular}[c]{@{}c@{}}Params. \\ (M)\end{tabular} &
    		\begin{tabular}[c]{@{}c@{}}FLOPs \\ (GMAC)\end{tabular} \\ \hline
    		1          & $27.296$          & $0.864$          & $0.259$          & $1.27$  & $17.55$  \\
    		2          & $28.091$          & $0.877$          & $0.206$          & $5.19$  & $26.18$  \\
    		\rowcolor{pink!50}\textbf{3} & $\mathbf{28.927}$ & $\mathbf{0.895}$ & $\mathbf{0.199}$ & $19.78$ & $35.27$  \\
    		4          & $27.876$ & $0.874$ & $0.222$ & $38.94$ & $82.59$  \\
    		5          & $27.671$& $0.871$ & $0.229$     & $46.11$ & $320.84$ \\ \hline
    	\end{tabular}
    	\label{table:ablationscale}
    \end{table}
    
    \subsubsection{Effect of Hybrid Unwinding Block}
\begin{figure} [htb]
    	\centering
    	\includegraphics[width=\linewidth]{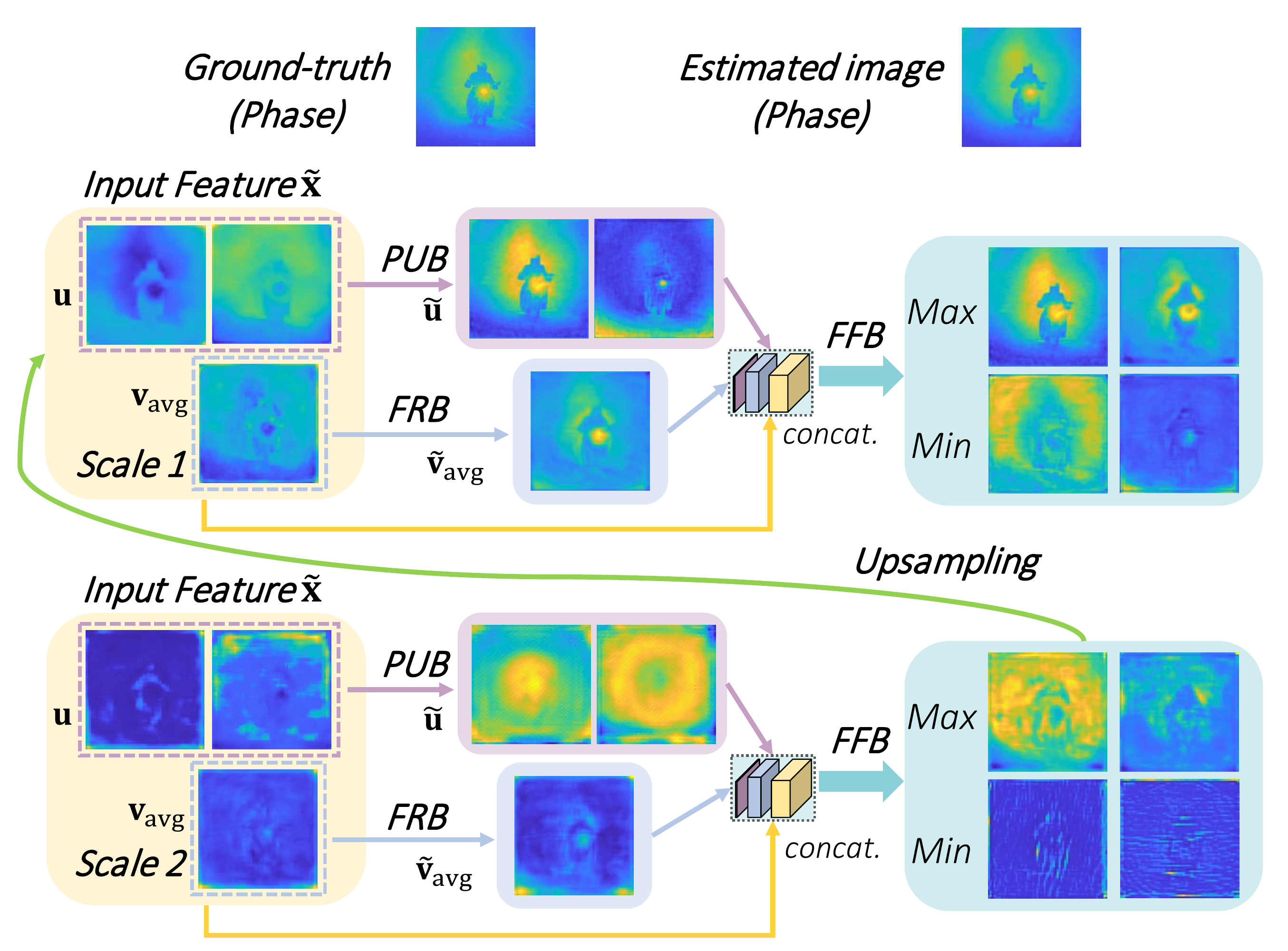}
    	\caption{Visualization of the feature maps generated by different components of HUB of the first two scales. The example image is from the COCO dataset.}
    	\label{fig:Vis_HUB}
    \end{figure}

    As introduced in Section \ref{sec:HUB}, HUB contains three components: PUB, FRB, and FFB. To gain a deeper insight into the operation of HUB, we visualized in Fig. \ref{fig:Vis_HUB} the feature maps generated by different components of the HUBs of the first two scales when inferencing an example image. The ground-truth and its final estimated image are shown at the top. At Scale $2$, the input feature maps are the upsampled FFB output of Scale $3$. The feature maps contain $256$ channels, with the first two fed to PUB and the rest fed to FRB. The outputs of PUB and FRB are shown. For better visualization of the FRB branch, the average feature maps of all channels are presented. It can be seen that PUB tends to give the global structure of the image, and FRB tends to give the details. These feature maps are then combined by FFB using the channel attention method. For FFB, we show the two output channels with the largest weights (denoted as \textit{Max}) and two with the smallest weights (denoted as \textit{Min}). The feature maps given by the channels with the largest weights have already contained the basic structure of the ground truth. They are sent to Scale $1$ for further enhancement. At Scale $1$, it can be seen the feature maps given by PUB are already very close to the ground truth. They are further processed to give the final output. To summarize, it can be seen that in Fig. \ref{fig:Vis_HUB} that the FFB outputs continuously improve from lower scale to upper scale. The physics information applied to the PUBs at different scales has played an important role in this enhancement process. It guides the network to reconstruct the image in the right direction not only at the training stage but also at the inferencing stage.  
    
    \subsubsection{Effect of Unwinding Layers}
    As presented in Section \ref{sec:HUB}, PUB is formed by $K$ unwinding layers. The feature maps are updated with the magnitude constraint $K$ times in the frequency domain. In order to determine this parameter $K$, we analyzed the performance with different numbers of unwinding layers in PUB. We considered seven different values of $K$, i.e., zero to six, in each PUB. TABLE \ref{table:ablationunwinding} presents the simulation results. It can be seen that using more layers (from zero to five) leads to better reconstruction performance (higher PSNR and SSIM, lower MAE). However, when the number of layers is larger than five, the performances tend to saturate. It is not surprising since, in deep learning, we have seen in many situations that deeper networks do not give better performance. Although using more layers allows more learnable parameters, it is also more difficult to train the network to give optimal performance. It can end up with the overfitting problem that unrelated details are introduced to fulfill the enlarged parameter space. Besides, the processing time also increases when more layers are used. Hence, we set the number of unwinding layers in PUB to five to obtain the best performance.
    
    \begin{table}[ht]
    	\centering
    	\caption{Comparison among different numbers of the unwinding layers on the RAF dataset (simulation). The best performances are marked in \textbf{bold} font. }
    	
    	\renewcommand{\arraystretch}{1.2}
    	\setlength\tabcolsep{7.9pt}
    	\begin{tabular}{cccccc}
    		\hline
    		\# Layers &
    		PSNR $\uparrow$ &
    		SSIM $\uparrow$ &
    		MAE $\downarrow$ &
    		\begin{tabular}[c]{@{}c@{}}Param.\\ (M)\end{tabular} &
    		\begin{tabular}[c]{@{}c@{}}FLOPs \\ (GMAC)\end{tabular} \\ \hline
    		0 & $15.369$ & $0.446$ & $1.019$ & $18.34$ & $24.37$\\
    		1 & $20.144$ & $0.632$ & $0.513$ & $18.64$ & $26.55$ \\
    		2 & $25.293$ & $0.816$ & $0.298$ & $18.93$ & $28.73$ \\
    		3 & $26.203$ & $0.838$ & $0.261$ & $19.21$ & $30.91$ \\
    		4 & $27.713$          & $0.874$          & $0.226$          & $19.49$ & $33.09$ \\
    		\rowcolor{pink!50} \textbf{5} & $\mathbf{28.927}$ & $\mathbf{0.895}$ & $\mathbf{0.199}$ & $19.78$ & $35.27$   \\ 
    		6 & $27.941$          & $0.874$          & $0.215$          & $20.06$ & $37.45$ \\ \hline
    	\end{tabular}
    	\label{table:ablationunwinding}
    \end{table}

    \section{Experiment On Optical Platform} \label{Sec:Experiment}
    
\begin{figure} [htb]
    	\centering
    	\includegraphics[width=\linewidth]{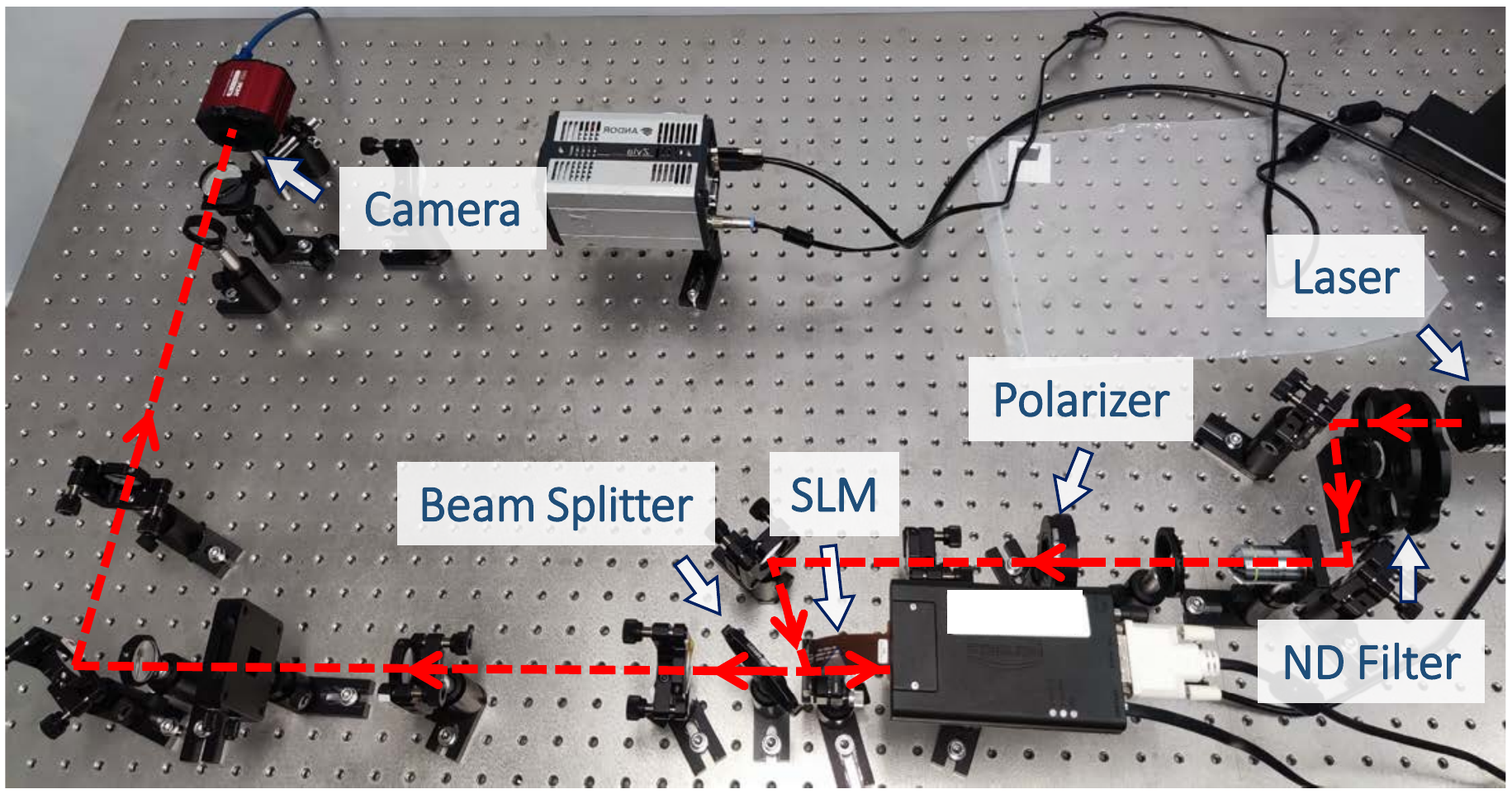}
    	\caption{The hardware implementation of the optical path in Fig. \ref{fig:opticalpath}.}
    	\label{fig:setup}
    \end{figure}

    To understand the performance of the proposed \textit{PPRNet} when applied to practical applications, we constructed an optical setup for collecting realistic intensity measurements for training and testing the network. The setup follows the optical path in Fig. \ref{fig:opticalpath}. A snapshot of it is shown in Fig. \ref{fig:setup}. It comprises a Thorlabs $ 10 mW$ HeNe laser with wavelength $ \lambda = 632.8 nm$ and a $ 12 $-bit $ 1920\times1200 $ Kiralux CMOS camera with a pixel pitch $ 5.86 \mu m$. We utilized a $ 1920\times1080 $ Holoeye Pluto phase-only SLM with pixel pitch $ \delta_{SLM} = 8\mu m $ to generate the target images. The SLM can impose different phase shifts to the coherent light gets through it. It effectively synthesizes the objects in phase imaging applications \cite{Sinha17, YE2022106808}. Since the ground truth is known, we can easily evaluate the accuracy of the reconstructed phase images given by different approaches. Another advantage of using the SLM is, similar to that in the simulation, we can directly multiple the defocus kernel with the images to generate the defocusing effect. 
    
    For the training and testing datasets, we used the same RAF and Fashion-MNIST datasets as in the simulation. Besides, we added the COCO dataset \cite{COCO_2014} (used the first $50000$ and $2000$ images of the training and testing dataset, respectively), which contains more complex images for challenging the PR approaches. We used the same approach as in the simulation to generate the phase part of the images based on the images collected from these datasets. The magnitude part was assumed as constant. Then these phase images were multiplied with the defocus kernel and loaded to the SLM. Following the optical path, the images captured by the camera were the defocused Fourier intensity measurements of these phase images. They were used for the training and testing of different PR approaches. The resolutions of these intensity measurements and phase images were the same as in the simulation, i.e., $762\times762$ and $128\times128$ pixels, respectively. When training the models, we first initialized the models with the ones we used in the simulation, as discussed in Section \ref{sec:simulation}. Then we fine-tuned all the pre-trained models by re-training them for $120$ epochs with the experimental measurements as the new training inputs. 
    
    We compared the proposed \textit{PPRNet} with four traditional optimization-based algorithms: GS \cite{Gerchberg1972APA}, HIO \cite{Fienup:82}, WF \cite{candes2015phase}, and RAAR \cite{Luke_2004}; and eleven deep learning-based methods: ResNet \cite{Nishizaki_2020, He_2016_CVPR}, PRCGAN \cite{uelwer2021phase}, LearnInitNet \cite{Morales_22}, $3$-scale UNet \cite{ronneberger2015u}, LenlessNet \cite{Sinha17}, CPR-FS \cite{Uelwer_PhaseRetrieval}, SiSPRNet \cite{Ye_SiSPRNet},  NNPhase \cite{Wu_cw5029}, and MCNN \cite{Wang_2020}, prDeep \cite{pmlr_metzler18a}, and HIO-UNet \cite{I_l_2019}. Similar to that in the simulation, we modified the input of some of the deep learning-based approaches since they originally could only handle much smaller size intensity measurements. Specifically, for CPR-FS, SiSPRNet, ResNet, NNPhase, MCNN, and PRCGAN, we inserted a pre-processing block before the original network structures. It contains two convolutional layers with a $5\times5$ kernel and strides $3$ and $4$, respectively, to extract the features and compress the dimension of the inputs from $762\times762$ to $128\times128$.  Besides, we also modified the networks that were originally designed to give real-valued images to reconstruct complex-valued images. Specifically, we set the output channels of LenslessNet, HIO-UNet, and ResNet as $2$ to give the real and imaginary parts of the images; and we also set the number of output channels of PRCGAN to $2$ for magnitude-phase representation.
    
    We evaluate the performance of all competing PR methods using the same metrics as in the simulation, i.e., PSNR, SSIM, and MAE. For the traditional optimization-based algorithms, we fine-tuned the hyper-parameters and fixed them when evaluating with the testing images. We did three trials of evaluation for each testing image, where each trial had $1500$ iterations. The reconstructed image with the highest PSNR among the three trials was chosen for comparison. Besides, traditional optimization-based algorithms can have trivial ambiguities in the reconstructed images, such as global phase shift, etc. They were removed before comparisons were conducted. 
    
\begin{figure*}[!ht]
    	
    	\centering
    	\begin{subfigure}{\textwidth}	
    		\centering
    		\begin{adjustbox}{width=\textwidth,center}       
    			\begin{tabular}{*{11}{c@{\extracolsep{0.1em}}} }
    				\includegraphics[width=0.1\textwidth,valign=t]{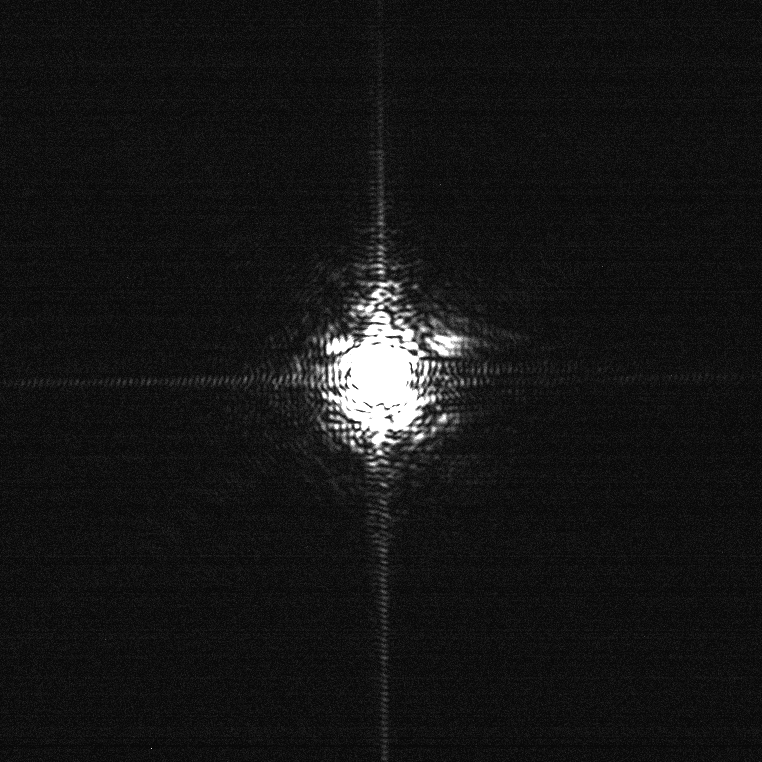}
    				&\includegraphics[width=0.1\textwidth,valign=t]{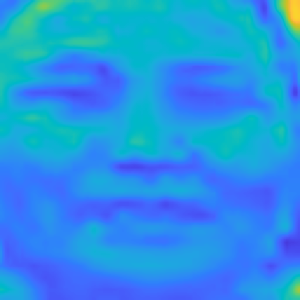}
    				&\includegraphics[width=0.1\textwidth,valign=t]{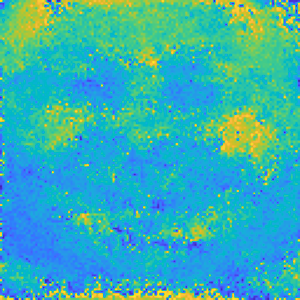}
    				&\includegraphics[width=0.1\textwidth,valign=t]{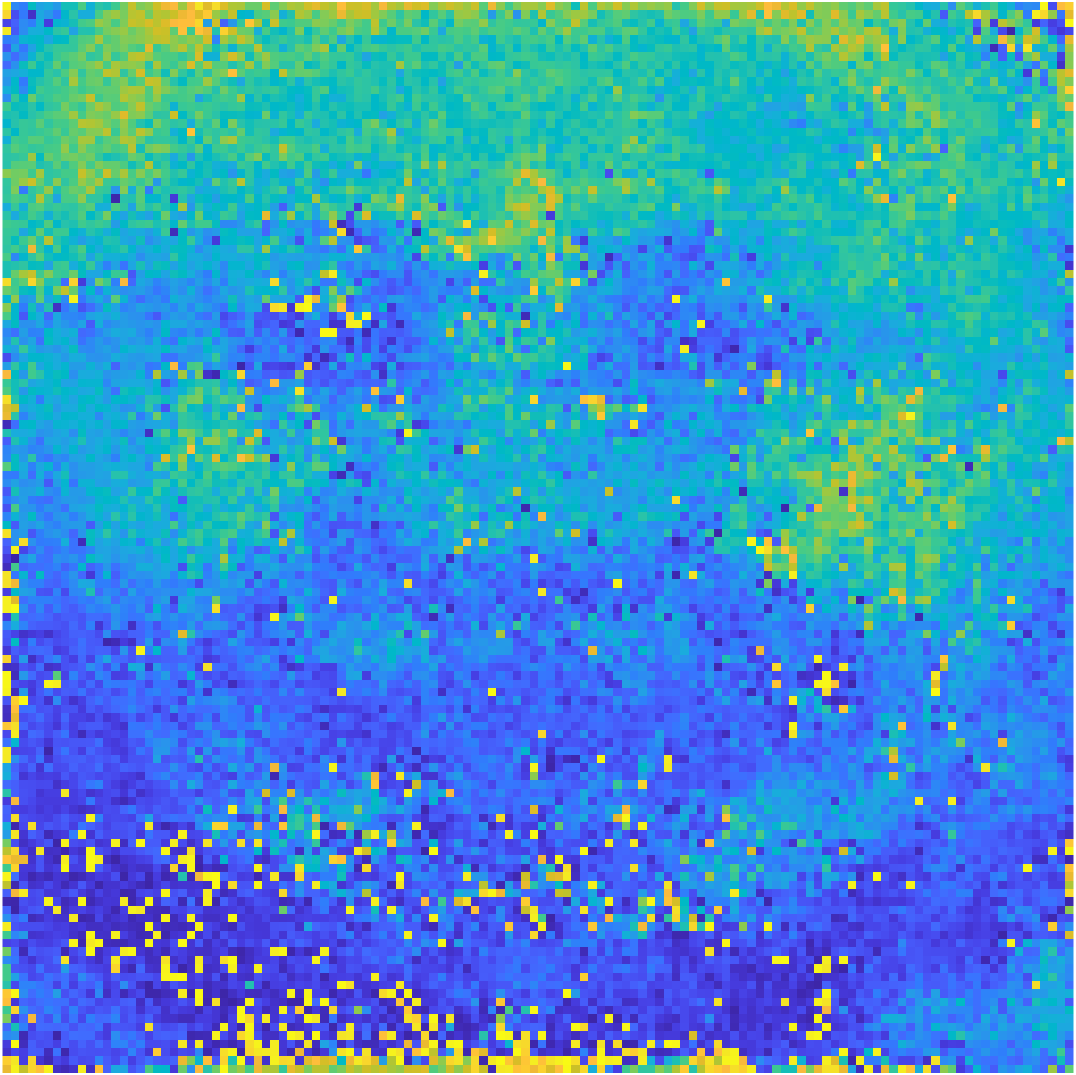}
    				&\includegraphics[width=0.1\textwidth,valign=t]{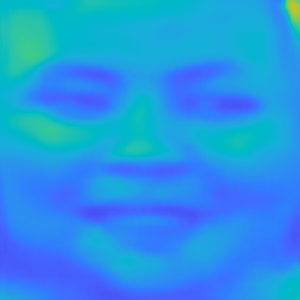}
    				&\includegraphics[width=0.1\textwidth,valign=t]{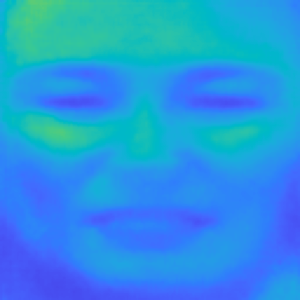}
    				&\includegraphics[width=0.1\textwidth,valign=t]{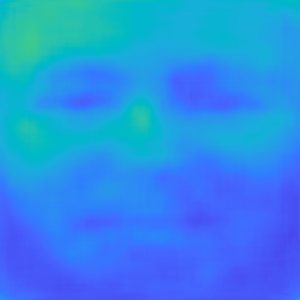}
    				&\includegraphics[width=0.1\textwidth,valign=t]{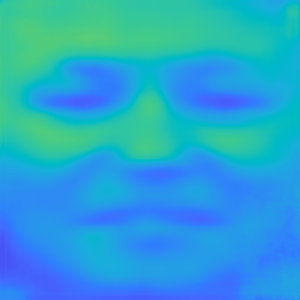}
    				&\includegraphics[width=0.1\textwidth,valign=t]{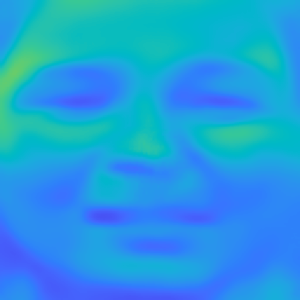}
    				&\includegraphics[width=0.1\textwidth,valign=t]{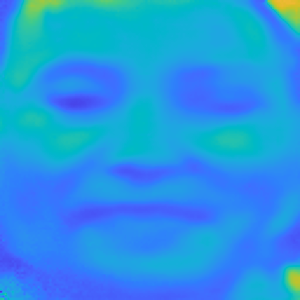}
    				&\includegraphics[width=0.1\textwidth,valign=t]{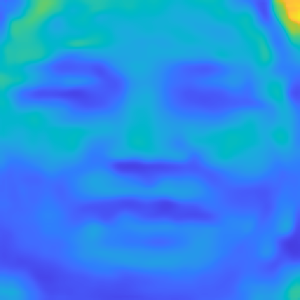}\\\addlinespace[0.5em]
    				
    				\includegraphics[width=0.1\textwidth,valign=t]{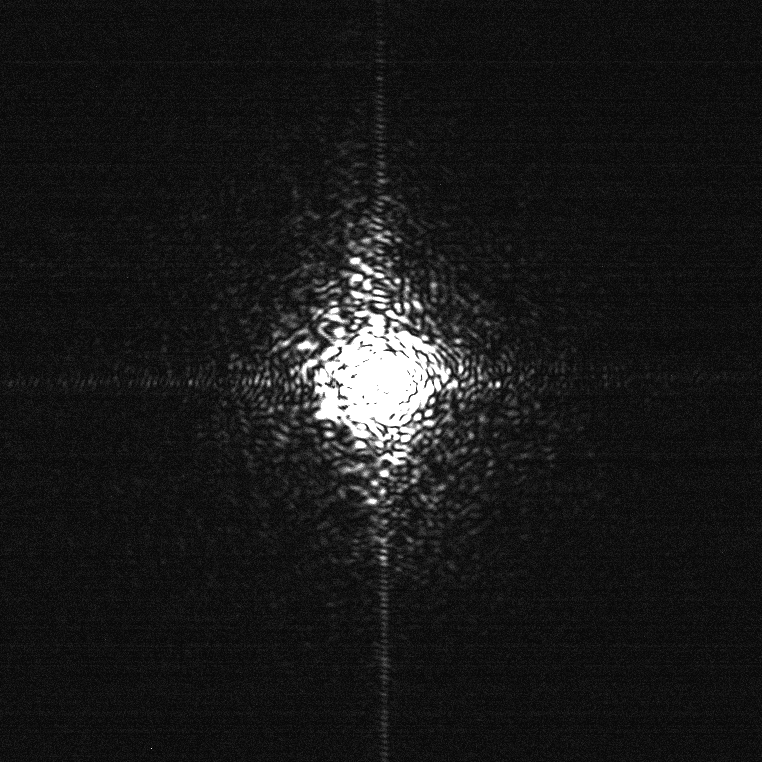}
    				&\includegraphics[height=0.118\textwidth,width=0.105\textwidth,valign=t]{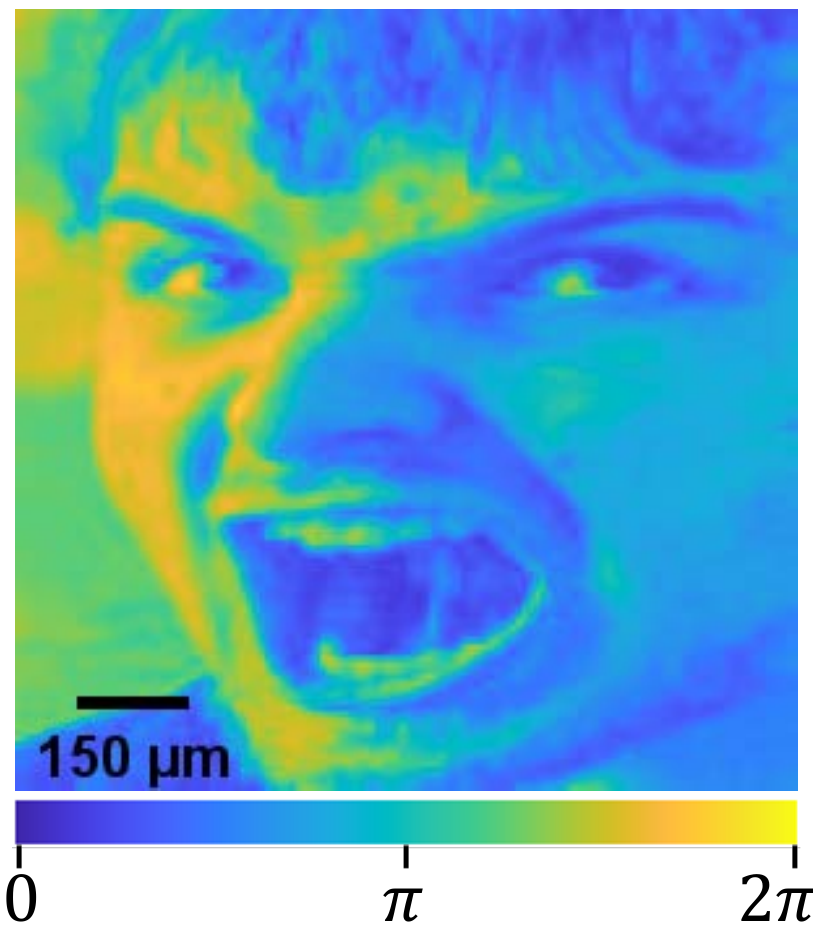}
    				&\includegraphics[width=0.1\textwidth,valign=t]{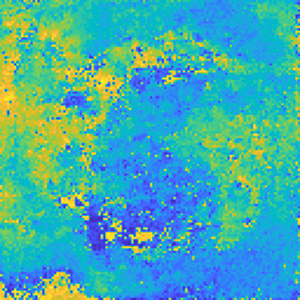}
    				&\includegraphics[width=0.1\textwidth,valign=t]{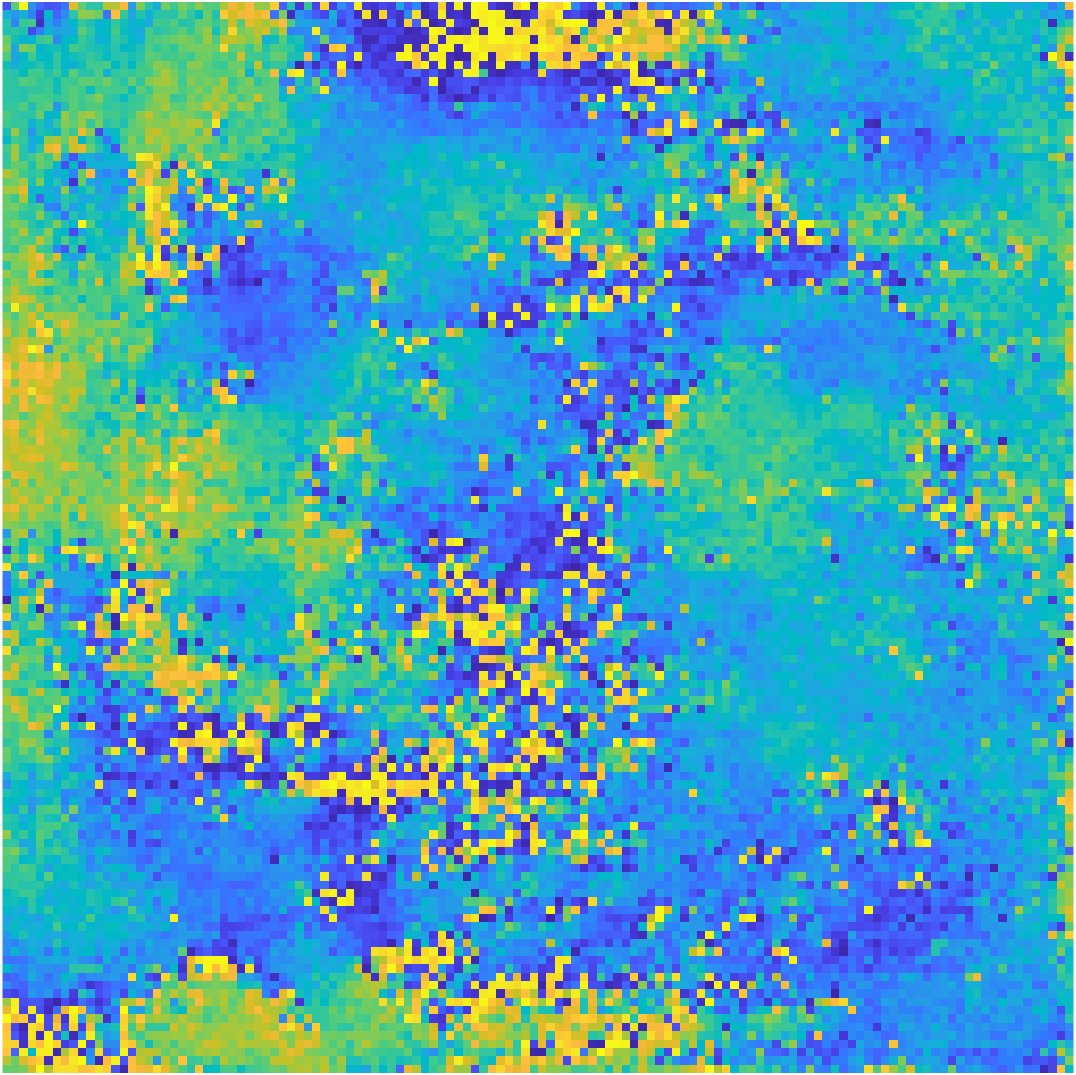}
    				&\includegraphics[width=0.1\textwidth,valign=t]{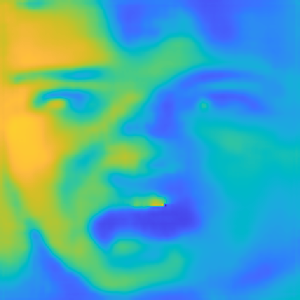}
    				&\includegraphics[width=0.1\textwidth,valign=t]{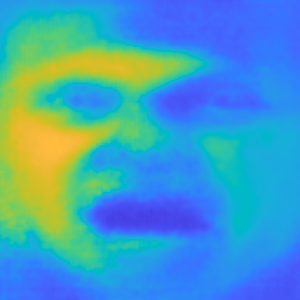}
    				&\includegraphics[width=0.1\textwidth,valign=t]{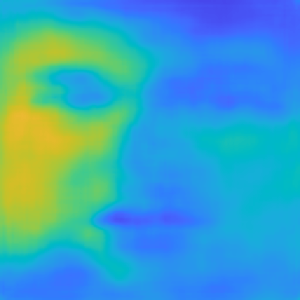}
    				&\includegraphics[width=0.1\textwidth,valign=t]{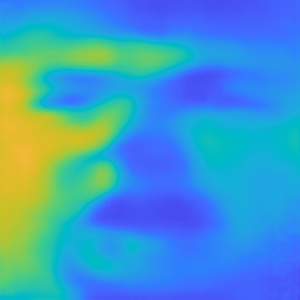}
    				&\includegraphics[width=0.1\textwidth,valign=t]{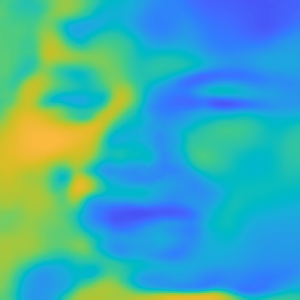}
    				&\includegraphics[width=0.1\textwidth,valign=t]{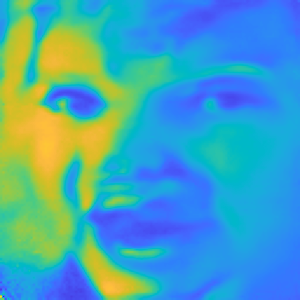}
    				&\includegraphics[width=0.1\textwidth,valign=t]{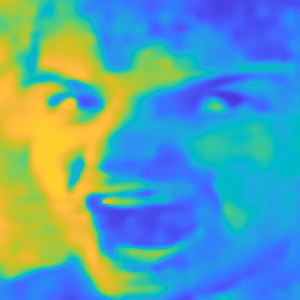}
    			\end{tabular}
    		\end{adjustbox}
    		\vspace*{-2.5mm}
    		\caption{}
    	\end{subfigure}
    	
    	\begin{subfigure}{\textwidth}	
    		\centering
    		
    		\begin{adjustbox}{width=\textwidth,center}       
    			\begin{tabular}{*{11}{c@{\extracolsep{0.1em}}} }
    				\includegraphics[width=0.1\textwidth,valign=t]{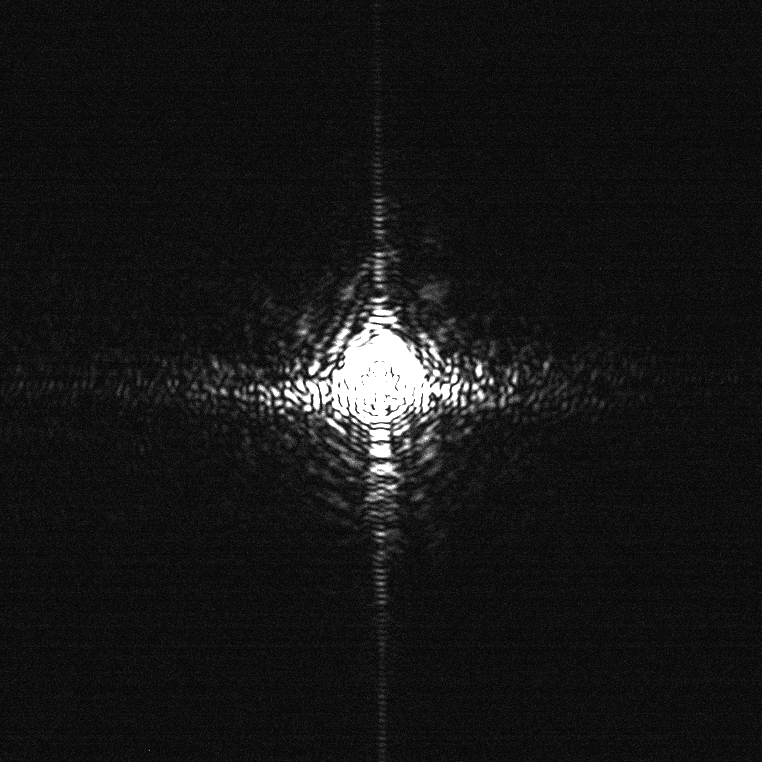}
    				&\includegraphics[width=0.1\textwidth,valign=t]{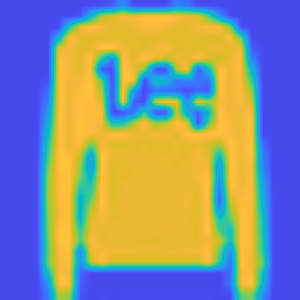}
    				&\includegraphics[width=0.1\textwidth,valign=t]{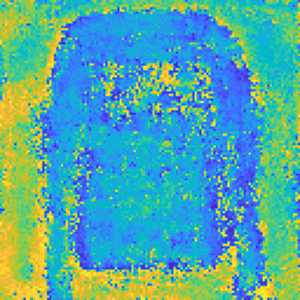}
    				&\includegraphics[width=0.1\textwidth,valign=t]{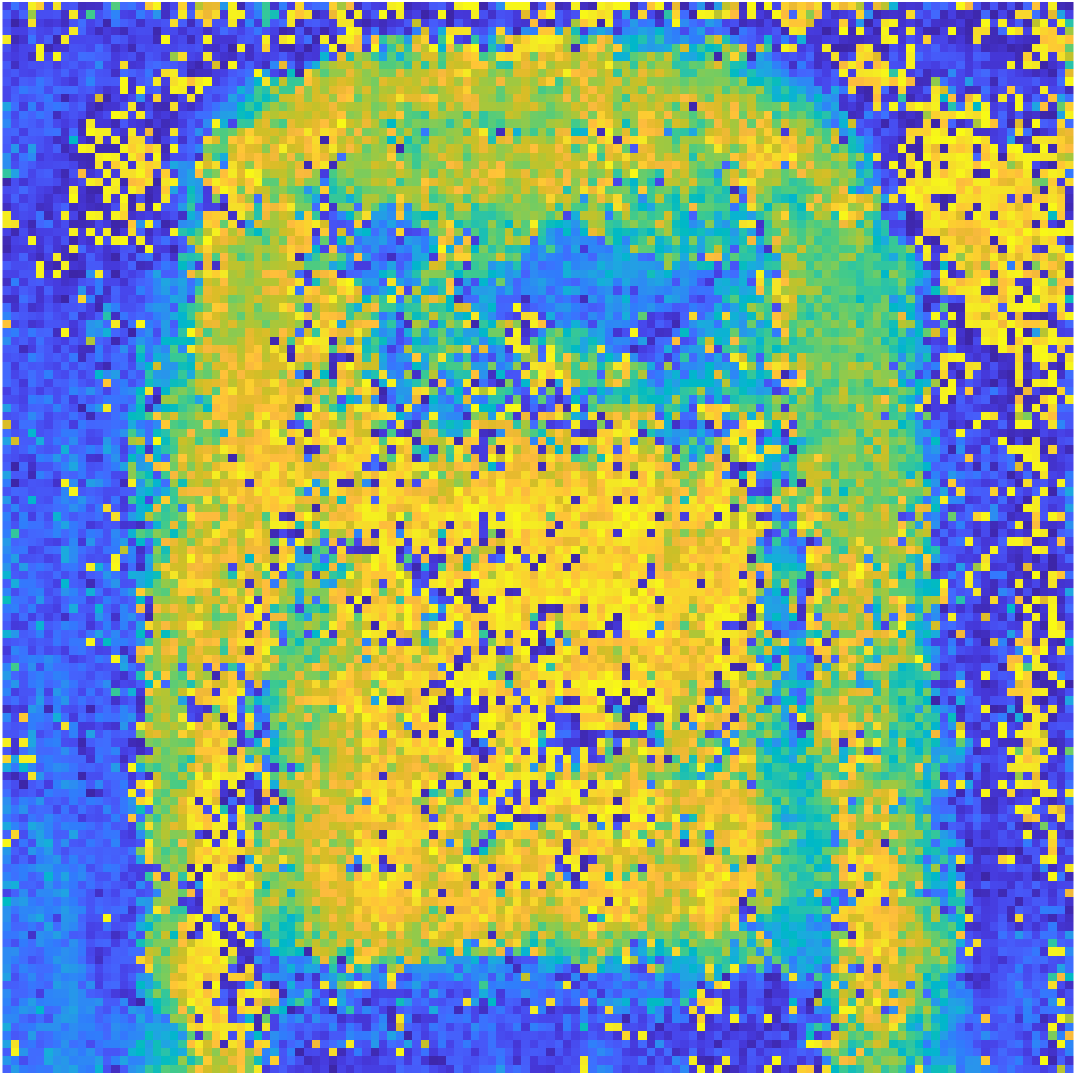}
    				&\includegraphics[width=0.1\textwidth,valign=t]{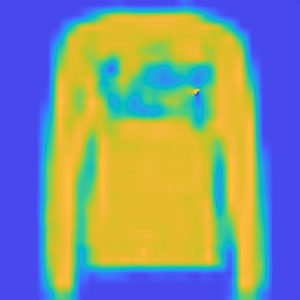}
    				&\includegraphics[width=0.1\textwidth,valign=t]{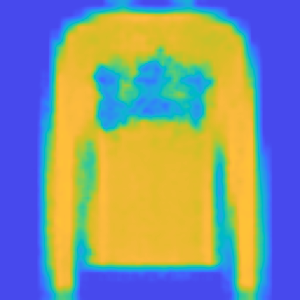}
    				&\includegraphics[width=0.1\textwidth,valign=t]{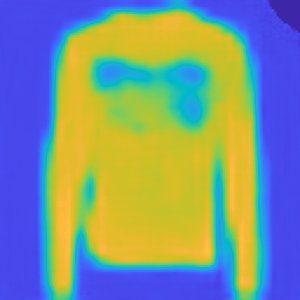}
    				&\includegraphics[width=0.1\textwidth,valign=t]{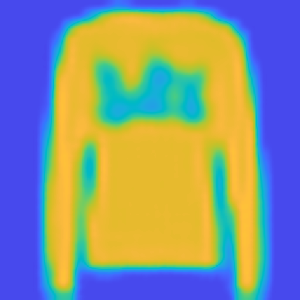}
    				&\includegraphics[width=0.1\textwidth,valign=t]{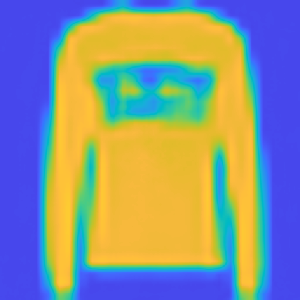}
    				&\includegraphics[width=0.1\textwidth,valign=t]{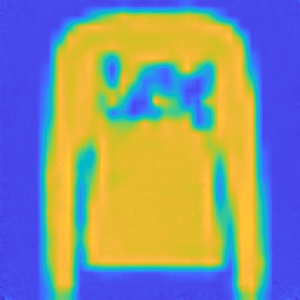}
    				&\includegraphics[width=0.1\textwidth,valign=t]{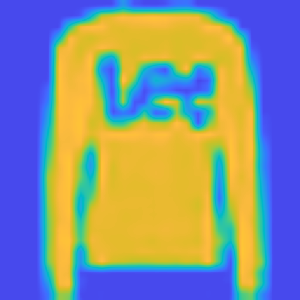}
    				\\\addlinespace[0.5em]
    				
    				\includegraphics[width=0.1\textwidth,valign=t]{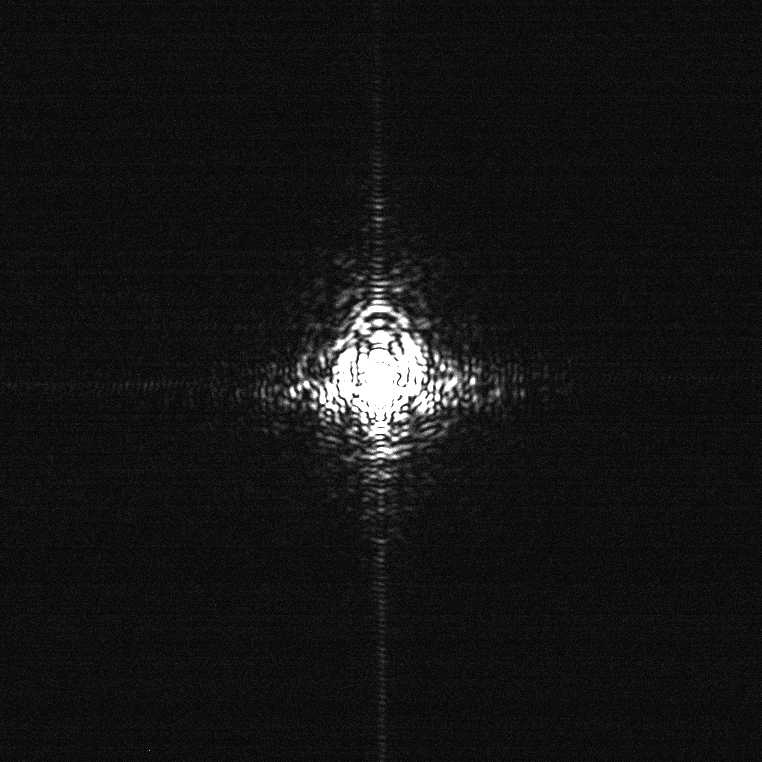}
    				&\includegraphics[height=0.118\textwidth,width=0.105\textwidth,valign=t]{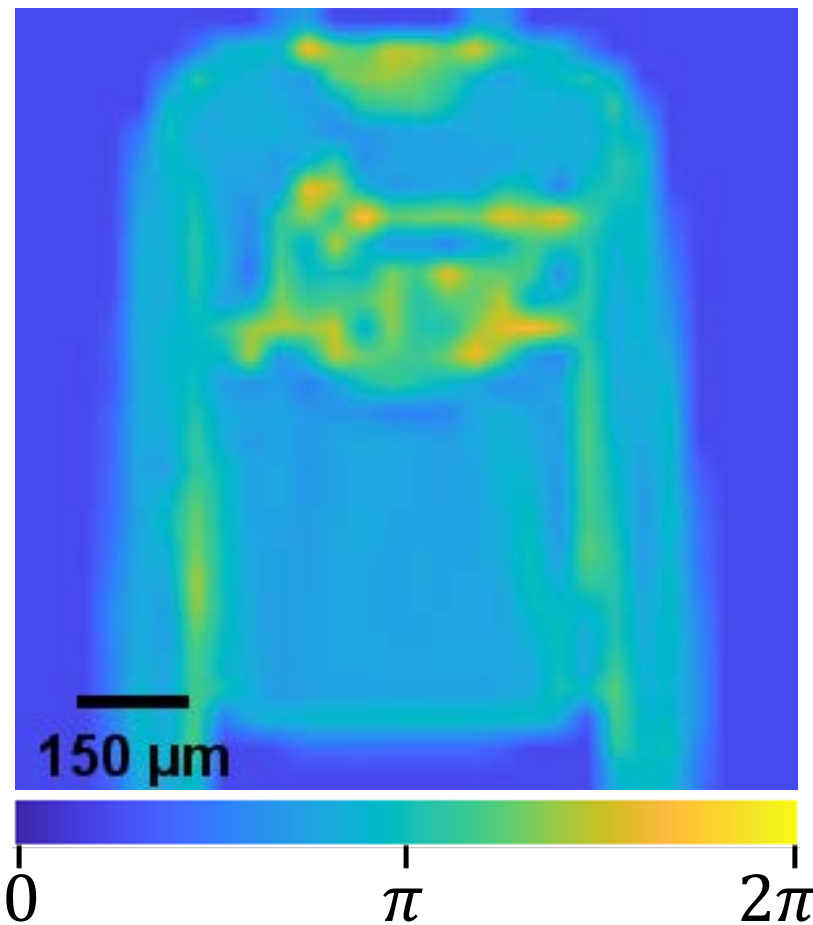}
    				&\includegraphics[width=0.1\textwidth,valign=t]{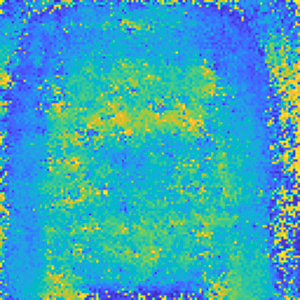}
    				&\includegraphics[width=0.1\textwidth,valign=t]{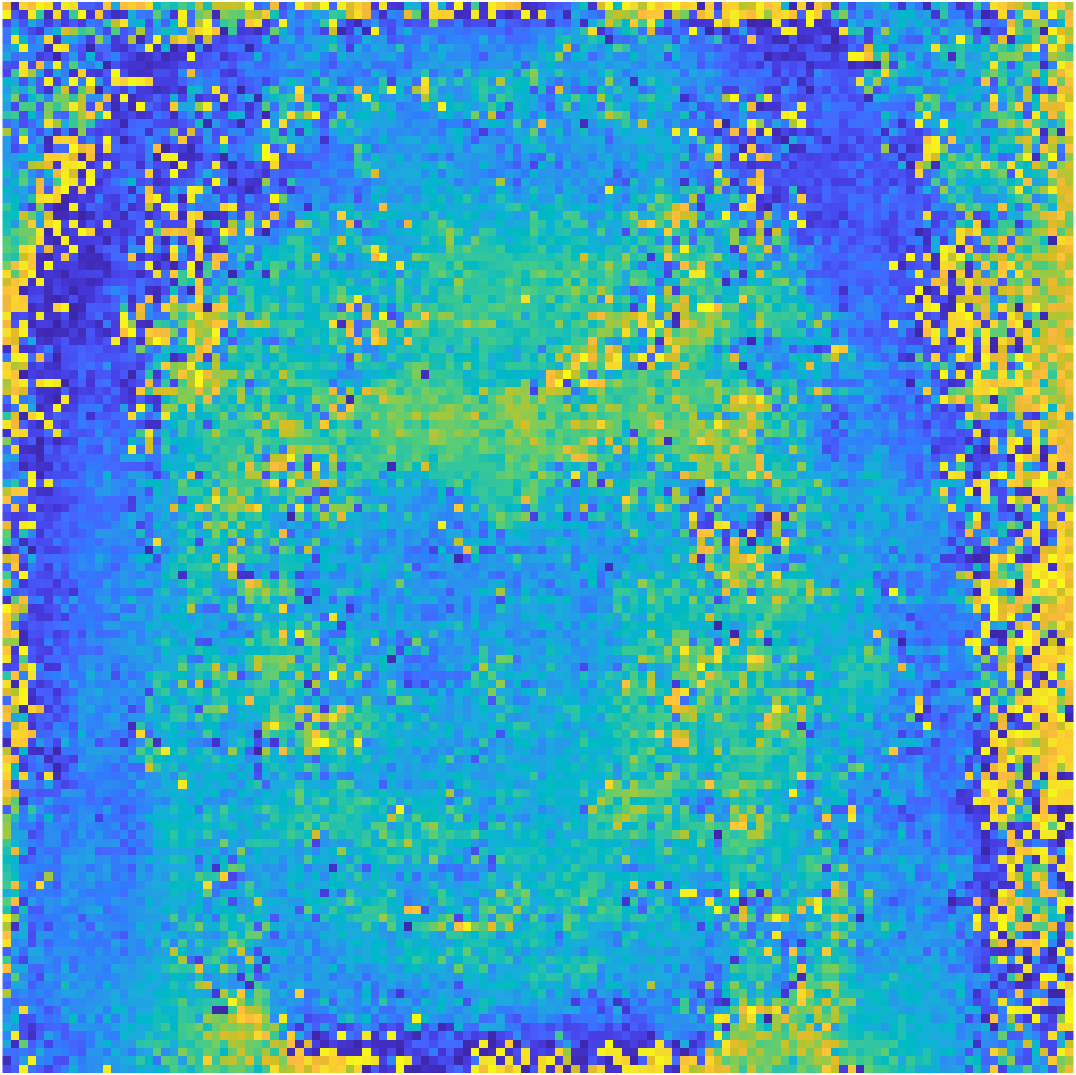}
    				&\includegraphics[width=0.1\textwidth,valign=t]{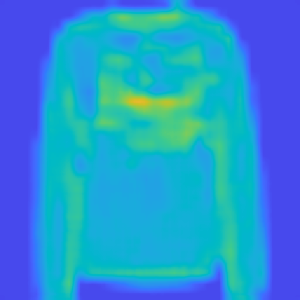}
    				&\includegraphics[width=0.1\textwidth,valign=t]{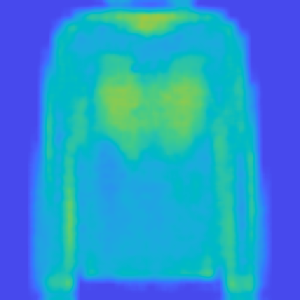}	
    				&\includegraphics[width=0.1\textwidth,valign=t]{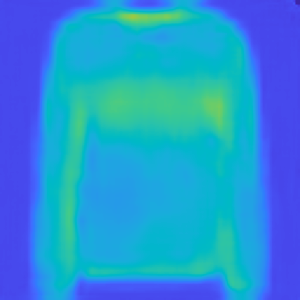}
    				&\includegraphics[width=0.1\textwidth,valign=t]{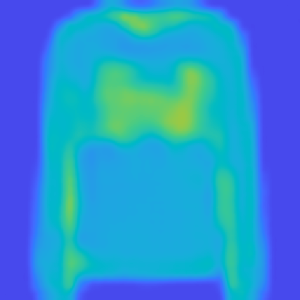}
    				&\includegraphics[width=0.1\textwidth,valign=t]{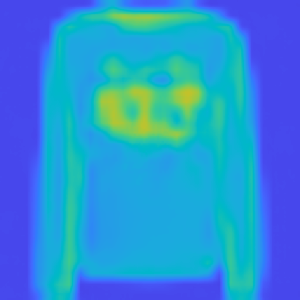}
    				&\includegraphics[width=0.1\textwidth,valign=t]{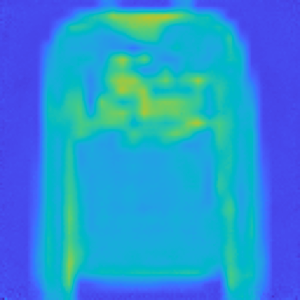}
    				&\includegraphics[width=0.1\textwidth,valign=t]{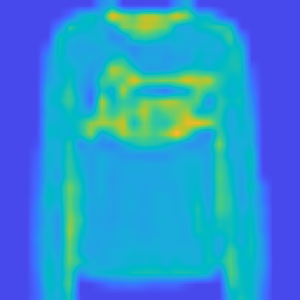}
    				\\
    			\end{tabular}
    		\end{adjustbox}
    		\vspace*{-2.5mm}
    		\caption{}
    	\end{subfigure}
    	
    	\begin{subfigure}{\textwidth}	
    		\centering
    		\begin{adjustbox}{width=\textwidth,center}       
    			\begin{tabular}{*{11}{c@{\extracolsep{0.1em}}} }
    				
    				\includegraphics[width=0.1\textwidth,valign=t]{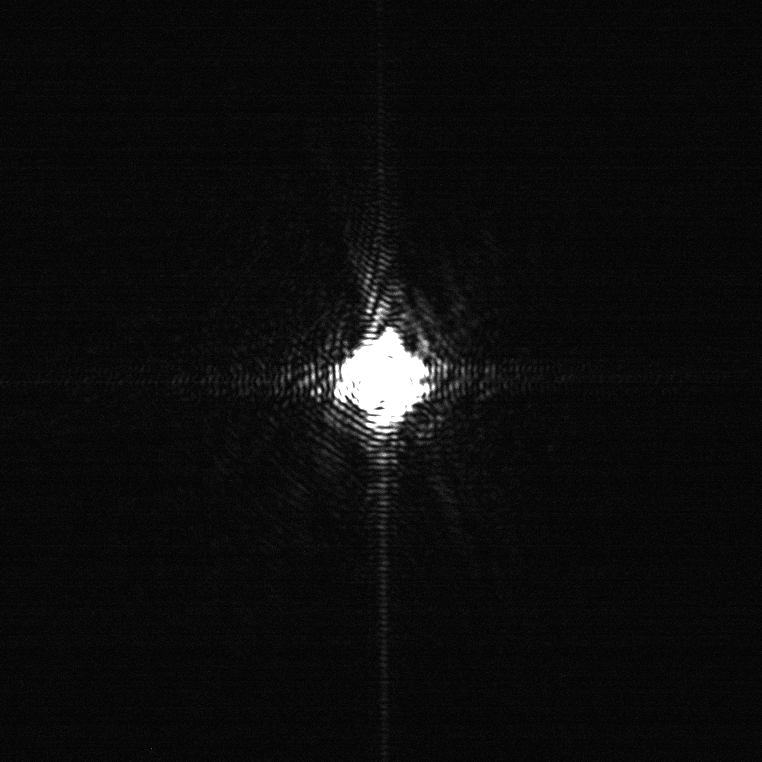}
    				&\includegraphics[width=0.1\textwidth,valign=t]{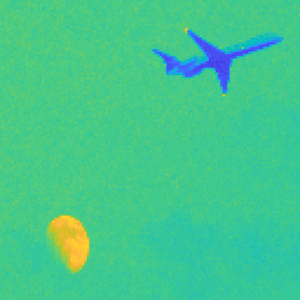}
    				&\includegraphics[width=0.1\textwidth,valign=t]{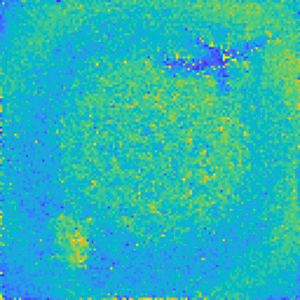}
    				&\includegraphics[width=0.1\textwidth,valign=t]{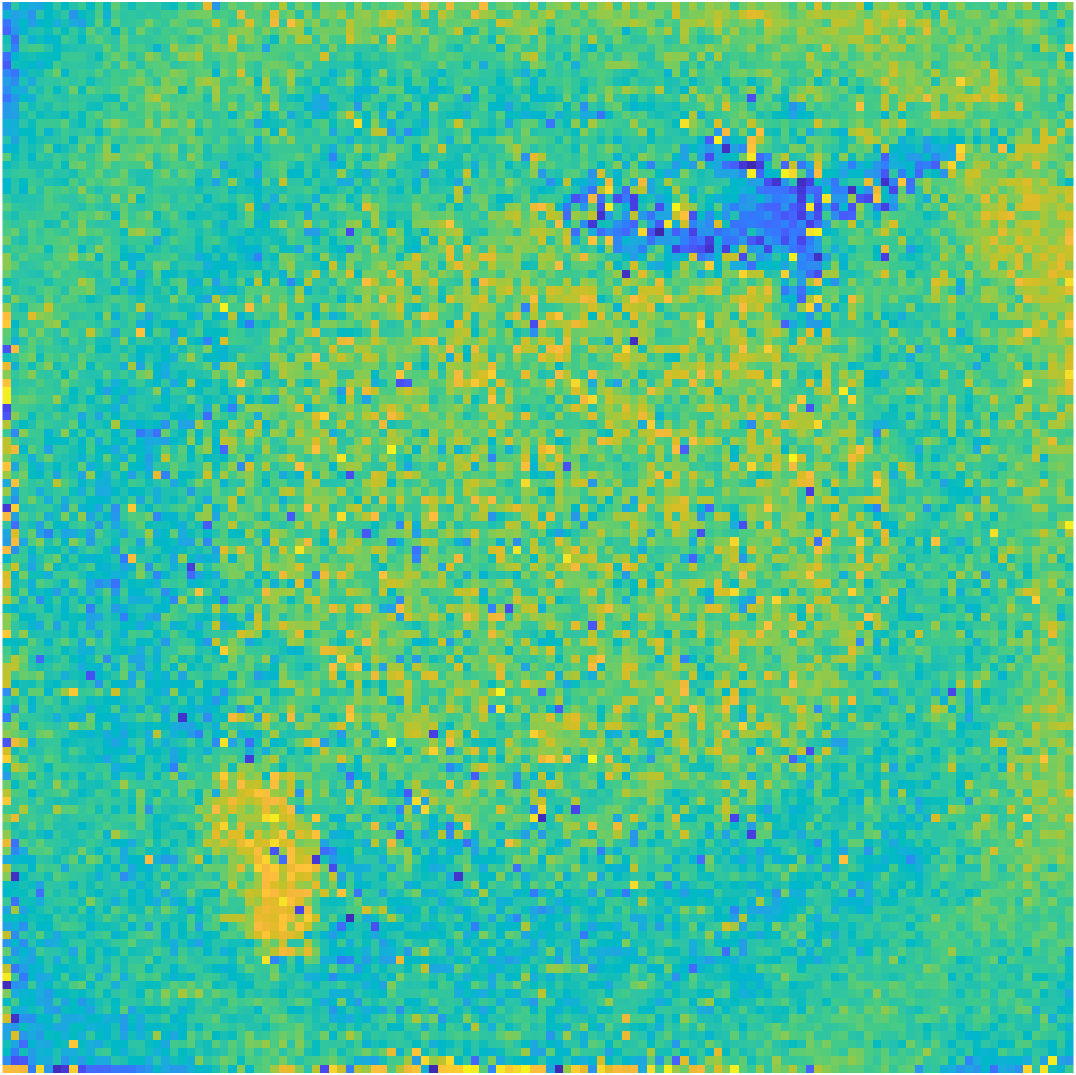}
    				&\includegraphics[width=0.1\textwidth,valign=t]{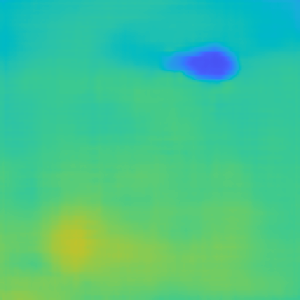}
    				&\includegraphics[width=0.1\textwidth,valign=t]{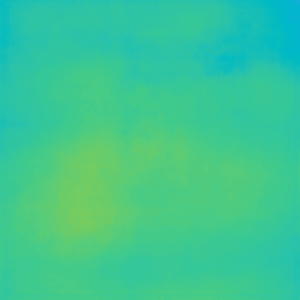}
    				&\includegraphics[width=0.1\textwidth,valign=t]{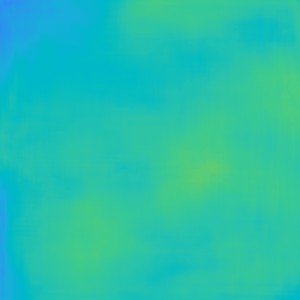}
    				&\includegraphics[width=0.1\textwidth,valign=t]{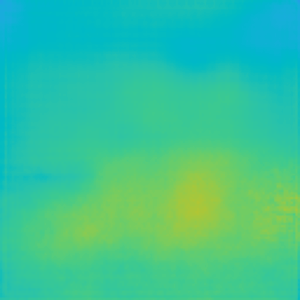}
    				&\includegraphics[width=0.1\textwidth,valign=t]{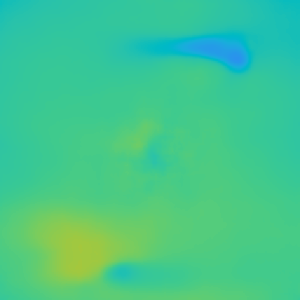}
    				&\includegraphics[width=0.1\textwidth,valign=t]{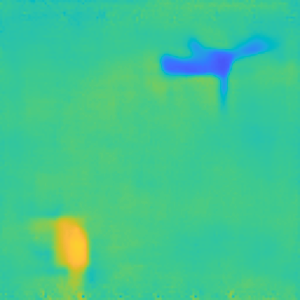}
    				&\includegraphics[width=0.1\textwidth,valign=t]{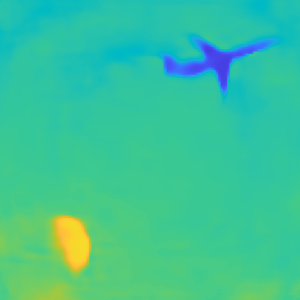}	\\\addlinespace[0.5em]
    				
    				\includegraphics[width=0.1\textwidth,valign=t]{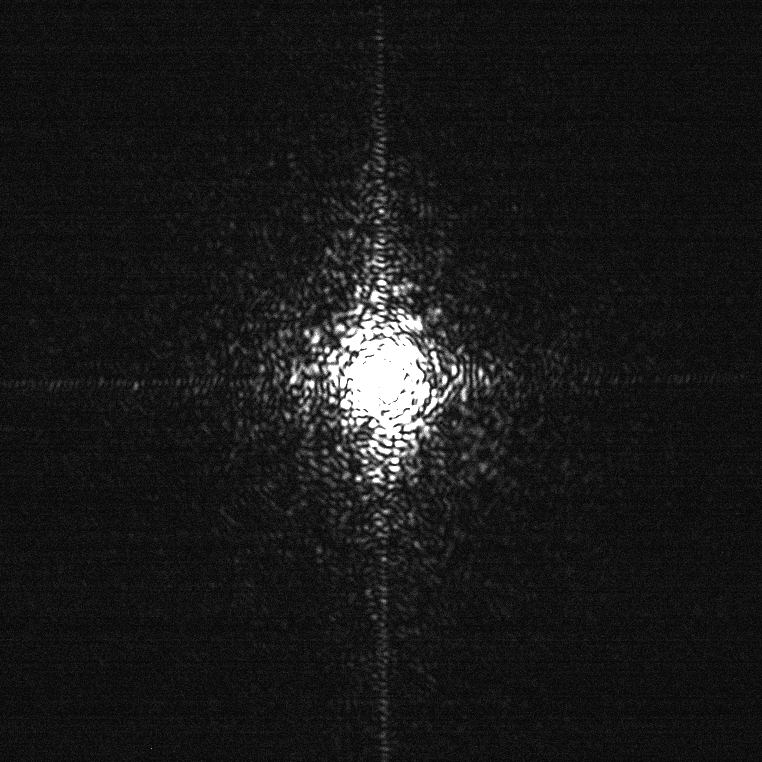}
    				&\includegraphics[width=0.1\textwidth,valign=t]{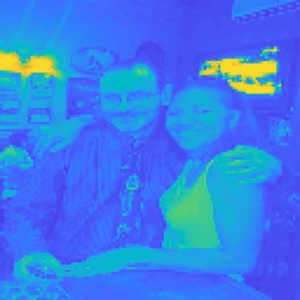}
    				&\includegraphics[width=0.1\textwidth,valign=t]{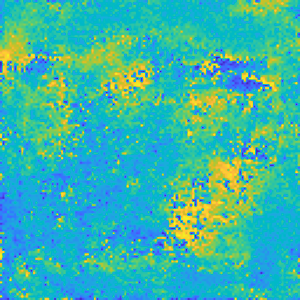}
    				&\includegraphics[width=0.1\textwidth,valign=t]{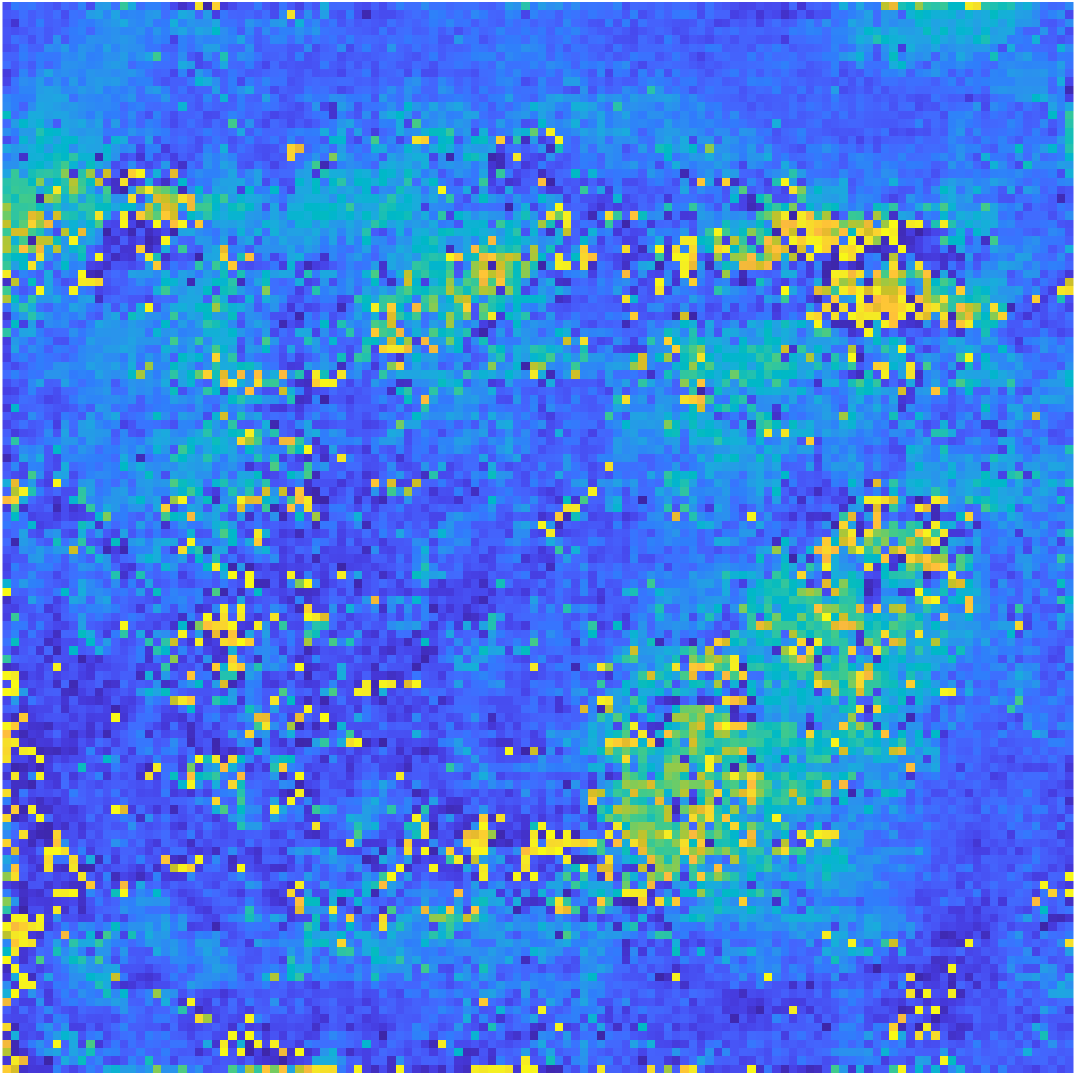}
    				&\includegraphics[width=0.1\textwidth,valign=t]{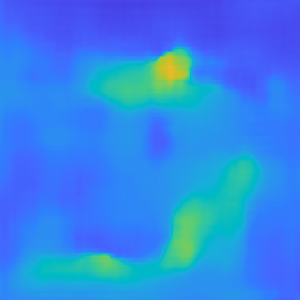}
    				&\includegraphics[width=0.1\textwidth,valign=t]{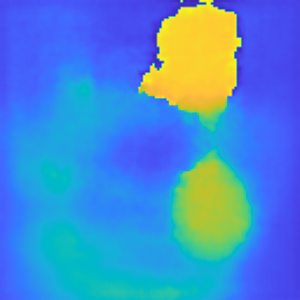}
    				&\includegraphics[width=0.1\textwidth,valign=t]{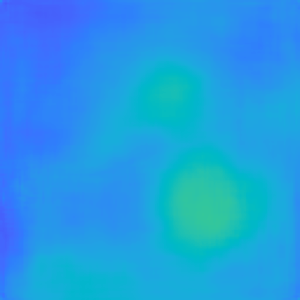}
    				&\includegraphics[width=0.1\textwidth,valign=t]{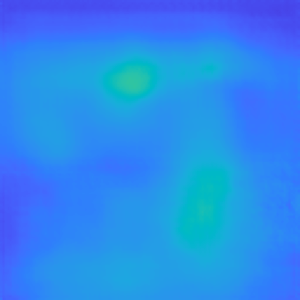}
    				&\includegraphics[width=0.1\textwidth,valign=t]{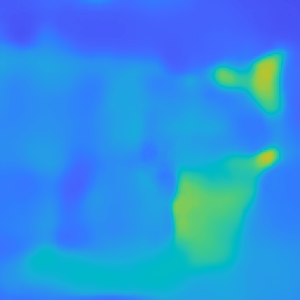}
    				&\includegraphics[width=0.1\textwidth,valign=t]{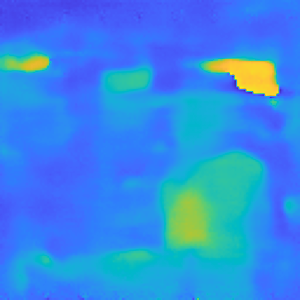}
    				&\includegraphics[width=0.1\textwidth,valign=t]{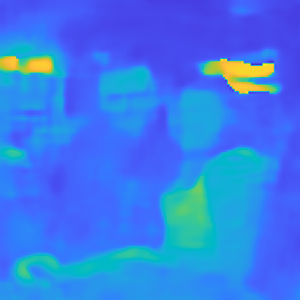}
    				\\\addlinespace[0.5em]
    				
    				\includegraphics[width=0.1\textwidth,valign=t]{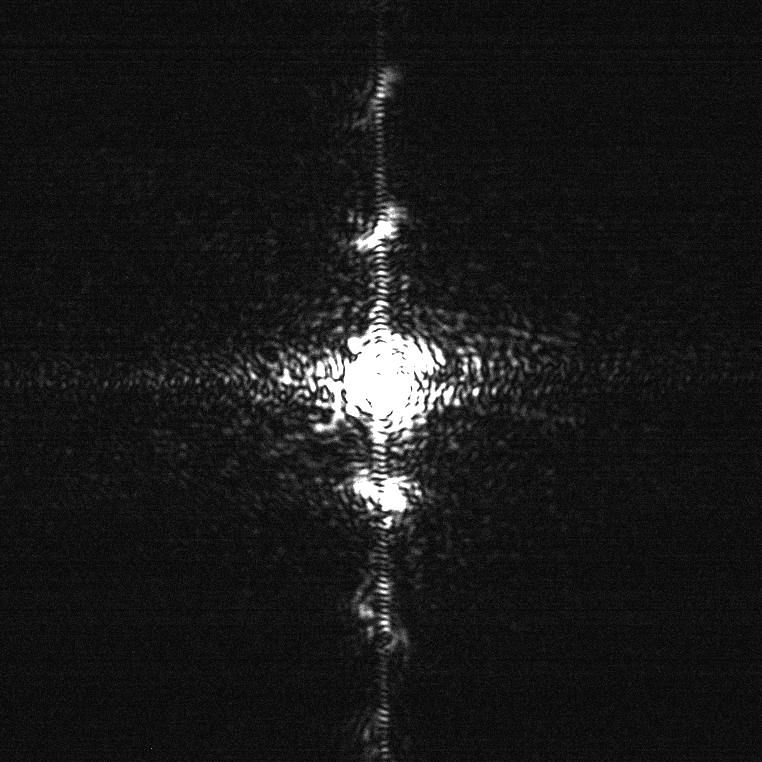}
    				&\includegraphics[height=0.118\textwidth,width=0.105\textwidth,valign=t]{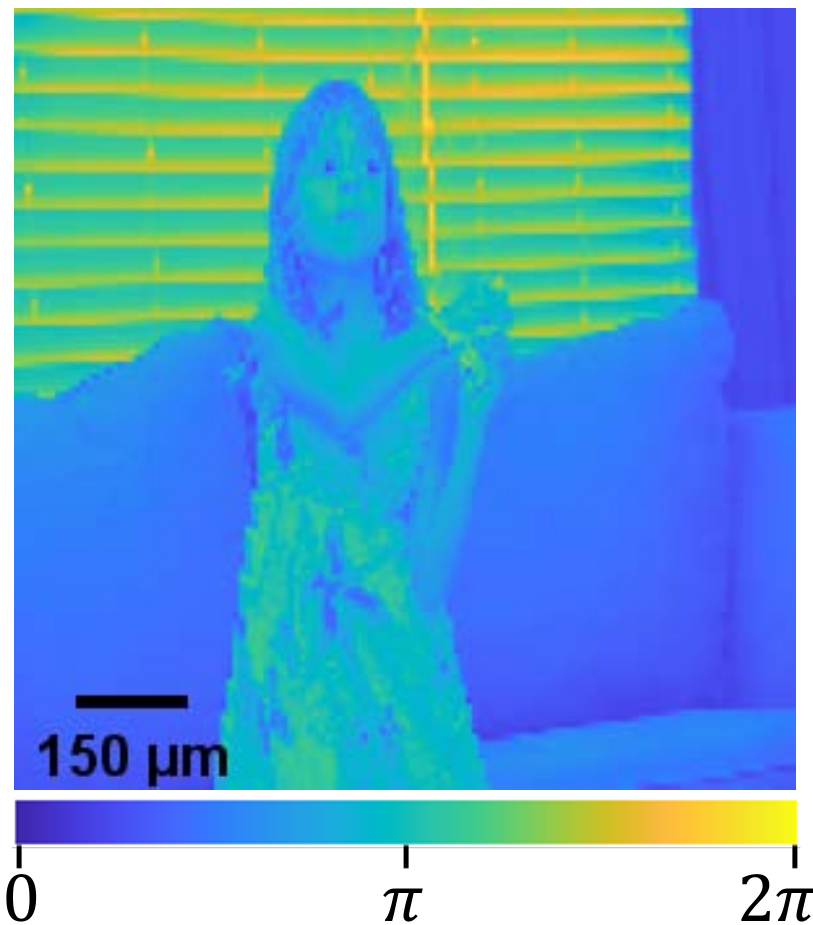}
    				&\includegraphics[width=0.1\textwidth,valign=t]{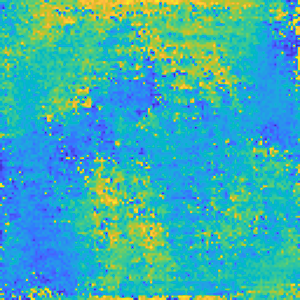}
    				&\includegraphics[width=0.1\textwidth,valign=t]{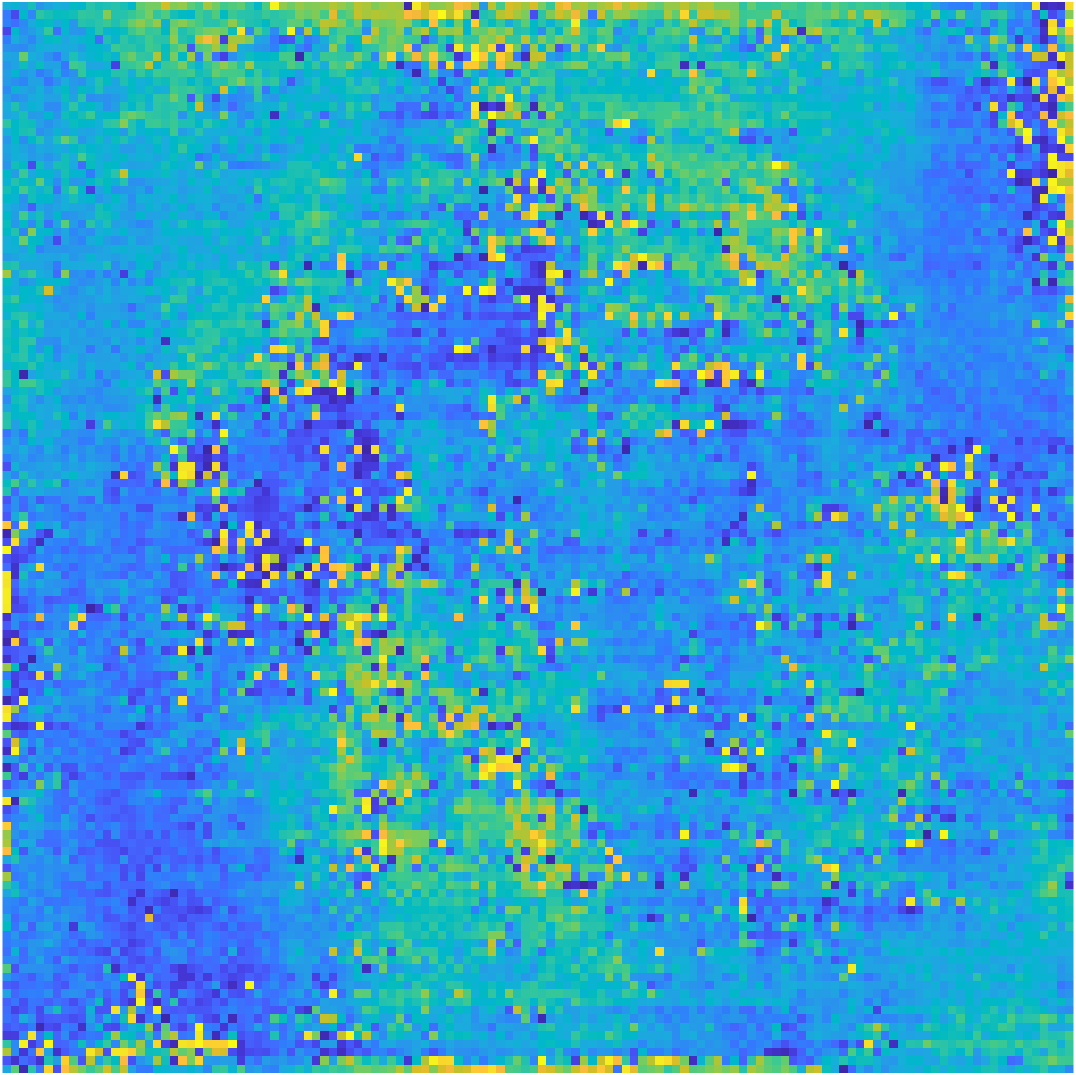}
    				&\includegraphics[width=0.1\textwidth,valign=t]{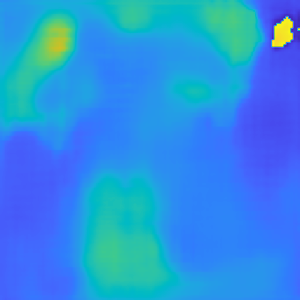}
    				&\includegraphics[width=0.1\textwidth,valign=t]{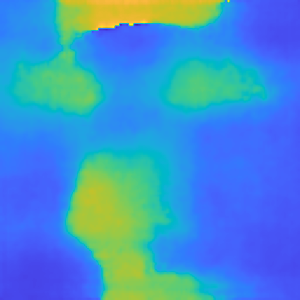}
    				&\includegraphics[width=0.1\textwidth,valign=t]{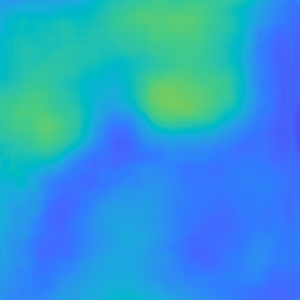}
    				&\includegraphics[width=0.1\textwidth,valign=t]{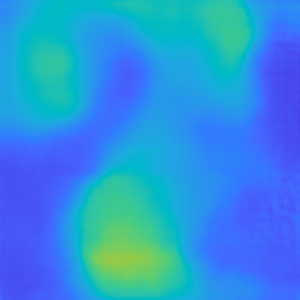}
    				&\includegraphics[width=0.1\textwidth,valign=t]{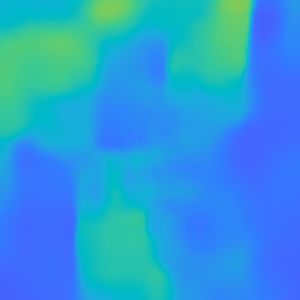}
    				&\includegraphics[width=0.1\textwidth,valign=t]{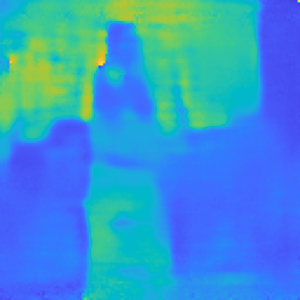}
    				&\includegraphics[width=0.1\textwidth,valign=t]{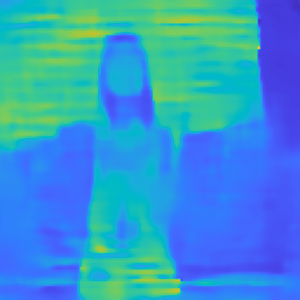}
    				\\
    				
    				{\scriptsize \makecell[c]{Fourier\\measurement}} & {\scriptsize \makecell[c]{Ground-truth \\ (phase)}} & {\scriptsize HIO } & {\scriptsize prDeep}
    				& {\scriptsize LenslessNet}  & {\scriptsize PRCGAN}  & {\scriptsize NNPhase} & {\scriptsize SiSPRNet} & {\scriptsize MCNN} & {\scriptsize HIO-UNet} & {\scriptsize \makecell[c]{\textbf{\textit{PPRNet}}\\(ours)} }
    				\\
    			\end{tabular}
    		\end{adjustbox}
    		\vspace*{-2mm}
    		\caption{}
    	\end{subfigure}
    	\vspace{-0.2cm}
    	\caption{Experimental results of different phase retrieval methods on (a) RAF dataset \cite{li2017reliable}, (b) Fashion-MNIST dataset \cite{xiao2017/online}, and (c) COCO dataset \cite{COCO_2014}. The first \textbf{column} and the second \textbf{column} show the Fourier intensity measurements (pixel values: $0 - 4095$) and the corresponding phase parts of ground-truth images with scale bars at the bottom left corners, respectively. The values of the Fourier measurements are scaled for better visualization. The other \textbf{columns} denote the reconstruction images through different methods. Except for the Fourier intensity measurements (the first column), the colormap of the rest columns ranges from $0$ to $2\pi$, with color bars at the second column.} 
    	\label{fig:comp_phaseonly}
    	
    \end{figure*}

    The quantitative and qualitative evaluation results are shown in TABLE \ref{table:comp_phaseonly} and Fig. \ref{fig:comp_phaseonly}, respectively. To save space, we do not include the qualitative results of GS, WF, RAAR, ResNet, UNet, LearnInitNet, and CPR-FS in Fig. \ref{fig:comp_phaseonly} due to their inferior performances. As presented in  TABLE \ref{table:comp_phaseonly}, the reconstruction performances of the proposed \textit{PPRNet} are much better than those of the compared approaches. Compared with the traditional optimization-based algorithms (GS, HIO, WF, and RAAR), \textit{PPRNet} has average PSNR and SSIM gains of at least $4.857dB$ and $0.266$, respectively, on the challenging COCO dataset that contains a wide range of rich content natural images. For the facial RAF dataset, the PSNR and SSIM gains can reach $12.219dB$ and $0.52$, respectively. The PSNR and SSIM gains can even reach $27.312dB$ and $0.704$, respectively, on the Fashion-MNIST dataset. In general, traditional optimization-based methods are difficult to converge to satisfactory results with only one intensity measurement. In particular, they have poorer performances in the Fashion-MNIST dataset. This is because the meaningful contents of the images (i.e., clothes) often located at the center while the values of outer regions are close to zeros. Thus, the algorithms fail to estimate the correct spatial support of the images. Besides, the saturated intensity measurement further complicates the problem, although it has already been lessened by using the defocusing method, as mentioned in Section \ref{sec:PRsystem}. As shown in Fig. \ref{fig:comp_phaseonly}, both HIO and RAAR methods can only recover the contour of the original images (third and fourth columns), but many defects remain. 
    
    Compared with the state-of-the-art deep learning-based PR methods, our \textit{PPRNet} can achieve PSNR and SSIM gains of at least $7.45dB$ and $0.08$, respectively, on the Fashion-MNIST dataset. It also performs the best on the RAF and COCO datasets. It can be seen in Fig. \ref{fig:comp_phaseonly} that the reconstructed images of the proposed \textit{PPRNet} can provide more details than other methods with fewer artifacts. It is particularly the case for the COCO dataset. Since the images are more complex, other approaches even cannot reconstruct the contours of the objects in the images. \textit{PPRNet} can recover not only the contours but provide more textures of the original images. It is benefited from the physics information of the underlying image that guides the reconstruction process. Compared with other physics-driven methods, such as HIO-UNet and PrDeep, \textit{PPRNet} not only provides better quality images but also has a lower complexity. It is expected since \textit{PPRNet} has a feedforward architecture that does not require iterative estimation. Besides, \textit{PPRNet} is end-to-end trained. It is different from HIO-UNet and PrDeep, which have an optimizer operated outside the network model. Thus, any mismatch between them can let the iteration process trap at the local minimum. Besides, prDeep is only successful for real-valued images, as claimed by the authors \cite{pmlr_metzler18a}. The result is unsatisfactory when it is used to retrieve phase images. The above experimental results have verified the performance of the proposed \textit{PPRNet} when used in a practical phase retrieval system. It consistently outperforms the state-of-the-art deep learning-based methods both quantitatively and qualitatively. 
    
\begin{table*}[!ht]
	\centering
	\caption{Quantitative comparison (average MAE/PSNR/SSIM/Parameters/Complexity)  with the state-of-the-art methods for phase retrieval on three datasets. The training and testing samples are collected with the optical system shown in Fig. \ref{fig:setup}. Best performances are marked in \textbf{bold} font. Second and third best performances are colored in {\color{blue} blue} and {\color{Green} green}, respectively.} 
	
			\begin{tabular}{cccc|ccc|ccc|c|c}
				\hline \multicolumn{1}{c}{\multirow{2}{*}{ Methods }}  & \multicolumn{3}{c|}{ Fashion-MNIST} & \multicolumn{3}{c|}{ RAF } & \multicolumn{3}{c|}{ COCO } & \multirow{2}{*}{\parbox{1.6cm}{\centering Parameters (Millions) }} & \multirow{2}{*}{\parbox{1.6cm}{\centering Complexity (GMAC) }}\\
				& PSNR $\uparrow$ & SSIM $\uparrow$ &  MAE $\downarrow$ & PSNR $\uparrow$ & SSIM $\uparrow$ & MAE $\downarrow$ & PSNR $\uparrow$ & SSIM $\uparrow$ &  MAE $\downarrow$  &  & \\
				\hline
				\multicolumn{12}{c}{Traditional Methods}\\ 
				\hline GS \cite{Gerchberg1972APA} & $9.178$ & $0.053$ & $1.681$ & $12.770$ & $0.129$ & $1.146$ & $10.985$ & $0.072$ & $1.421$ & NA & NA\\
				HIO \cite{Fienup:82} & $9.082$ & $0.052$ & $1.709$ & $12.553$ & $0.127$ & $1.185$ & $10.820$ & $0.069$ & $1.460$ & NA & NA\\
				WF \cite{candes2015phase} & $10.171$ & $0.271$ & $1.443$ & $14.254$ & $0.325$ & $0.908$ & $12.671$ & $0.191$ & $1.140$ & NA & NA\\
				RAAR \cite{Luke_2004} & $9.189$ & $0.054$ & $1.678$ & $12.783$ & $0.129$ & $1.143$  & $10.996$ & $0.072$& $1.418$ & NA & NA \\ 
				\hline \multicolumn{12}{c}{Learning-based Methods}\\ 
				\hline ResNet \cite{Nishizaki_2020} & $26.167$ & $0.839$ & $0.171$ & $20.322$ & $0.605$ & $0.461$  &  $14.998$ & $0.292$&  $0.872$ & $4.72$ & $77.47$\\
				UNet \cite{ronneberger2015u} & $20.201$ & $0.736$ &  $0.343$ & $17.309$ & $0.527$ & $0.688$ & $13.894$ & $0.267$& $1.062$ & $21.43$ & $27.09$\\
				LenslessNet \cite{Sinha17} &  $27.152$ &  $0.865$ &  $0.160$ & {\color{Green} $20.979$} &  $0.645$ & {\color{Green} $0.433$} & $14.810$ & {\color{Green}$0.295$}& $0.898$ & $32.74$ & $4.01$\\
				SiSPRNet \cite{Ye_SiSPRNet} & {\color{Green}$28.682$} & {\color{Green}$0.877$} &  {\color{Green} $0.141$} & $20.347$ & {\color{Green}$0.667$} & $0.458$ & $14.443$ & $0.277$ & $0.980$  & $19.51$ & $1.78$\\
				MCNN \cite{Wang_2020} & $27.594$ & $0.862$ &  $0.155$ & $20.611$ & $0.619$ & $0.457$ & {\color{Green}$15.170$} & $0.283$& $0.887$ & $28.55$ & $52.40$ \\
				PRCGAN \cite{uelwer2021phase} & $26.619$ & $0.846$ &  $0.167$ & $20.150$ & $0.603$ & $0.474$ & $13.971$ & $0.267$ & $0.991$& $111.38$ & $3.45$\\
				NNPhase \cite{Wu_cw5029} & $24.637$ & $0.807$ &  $0.222$ & $19.315$ & $0.567$ & $0.539$ & $14.598$ & $0.272$& $0.978$ & $18.85$ & $6.49$\\
				LearnInitNet \cite{Morales_22} & $23.771$ & $0.790$ &  $0.241$ & $19.024$ & $0.562$ & $0.560$ & $14.379$ & $0.271$& $0.996$ & $7.43$ & $1.27$\\
				CPR-FS \cite{Uelwer_PhaseRetrieval} & $18.965$ & $0.770$ &  $0.536$ & $16.426$ & $0.528$ & $0.767$ & $13.148$ & $0.261$& $1.204$ & $895.32$ & $0.12$\\
				prDeep \cite{pmlr_metzler18a} & $9.422$ & $0.057$ &  $1.478$ & $12.993$ & $0.130$ & $1.053$ & $12.274$ & $0.092$& $1.115$ & $0.56$ & $1828.01$\\
				HIO-UNet \cite{I_l_2019} & {\color{blue} $30.033$} & {\color{blue} $0.896$} &  {\color{blue} $0.132$} & {\color{blue} $22.527$} & {\color{blue} $0.701$} & {\color{blue} $0.366$} & {\color{blue} $16.314$} & {\color{blue} $0.346$} & {\color{blue} $0.748$} & $62.16$ & $142.87$\\
				\rowcolor{pink!50}\textbf{\textit{PPRNet} (Ours)} & $\mathbf{37.483}$ & $\mathbf{0.975}$ & $\mathbf{0.052}$ & $\mathbf{26.473}$ & $\mathbf{0.845}$ & $\mathbf{0.252}$   & $\mathbf{17.528}$ & $\mathbf{0.457}$& $\mathbf{0.671}$ & $19.78$ & $35.27$\\
				\hline
			\end{tabular}
		\label{table:comp_phaseonly}
	\end{table*}
	
\section{Conclusion} \label{Sec:Conclusion}
In this paper, we proposed a deep learning-based phase retrieval (PR) method, namely \textit{PPRNet}. Similar to other deep learning-based approaches, \textit{PPRNet} requires only a single Fourier intensity measurement as its input and does not need an extra masking scheme for the PR system to constrain the measurement data. The main novelty of \textit{PPRNet} is the introduction of physics information to the PR process. Unlike the traditional physics-driven PR methods that often end up in a time-consuming iterative procedure, the proposed \textit{PPRNet} has a non-iterative feedforward structure but can still effectively utilize the intensity measurement to guide the image reconstruction process. It is enabled by the novel Hybrid Unwinding Blocks (HUB) embedded in the network’s input and expanding path. It separately processes the global and local information of the feature maps with the aid of the intensity measurement and combines them with a channel attention method. Our simulation and experiment results have verified the effectiveness of \textit{PPRNet}. In particular, our experiment results were obtained from an optical platform designed for this research. They demonstrate the performance of \textit{PPRNet} when applied to practical phase retrieval applications. From the simulation and experiment results, it can be concluded that the proposed \textit{PPRNet} consistently outperforms the state-of-the-art deep learning-based PR methods, proving it a promising solution to practical PR applications. Nevertheless, at the moment, there is still room for \textit{PPRNet} to improve further when handling images with complex scenes. It is one of the ongoing researches in our group.

    \bibliographystyle{IEEEtran}
    \bibliography{Reference_All}

\end{document}